\documentclass[fleqn,usenatbib]{mnras}

\usepackage{mathptmx}

\usepackage[T1]{fontenc}

\DeclareRobustCommand{\VAN}[3]{#2}
\let\VANthebibliography\thebibliography
\def\thebibliography{\DeclareRobustCommand{\VAN}[3]{##3}\VANthebibliography}

\usepackage{graphicx}	
\usepackage{amsmath}	
\usepackage{amssymb}	
\usepackage{hyperref}
\usepackage{xcolor}

\title[Lensing masses of a new sample of galaxy groups]{On the weak lensing masses  of a new sample of galaxy groups}

\author[E. J. Gonzalez et al.]{
Elizabeth J. Gonzalez,$^{1,2,3}$\thanks{E-mail: ejgonzalez@unc.edu.ar}
Facundo Rodriguez,$^{2,3}$
Manuel Merch\'an,$^{2,3}$
Diego Garc\'ia Lambas,$^{2,3}$
\newauthor
Mart\'in Makler,$^{1,4}$
Mart\'in Chalela,$^{2,3}$
Maria E. S. Pereira,$^{5}$
Bruno Moraes,$^{6}$
and HuanYuan Shan$^{7,8}$
\\
$^{1}$Centro Brasileiro de Pesquisas F\'{\i}sicas, Rio de Janeiro, RJ 22290-180, Brazil\\
$^{2}$Instituto de Astronom\'{\i}a Te\'orica y Experimental (IATE-CONICET),
 Laprida 854, X5000BGR, C\'ordoba, Argentina\\
$^{3}$Observatorio Astron\'omico de C\'ordoba, Universidad Nacional de C\'ordoba, Laprida 854, X5000BGR, C\'ordoba, Argentina.\\
$^{4}$International Center for Advanced Studies \& ICIFI (CONICET), ECyT-UNSAM, Campus Miguelete, 25 de Mayo y Francia, CP1650, Buenos Aires, Argentina \\
$^{5}$Department  of  Physics,  University  of  Michigan,  Ann  Arbor,  MI  48109,  USA\\
$^{6}$Instituto de F\'{i}sica, Universidade Federal do Rio de Janeiro, 21941-972, Rio de Janeiro, RJ, Brazil\\
$^{7}$Shanghai Astronomical Observatory (SHAO), Nandan Road 80, Shanghai 200030, China\\
$^{8}$University of Chinese Academy of Sciences, Beijing 100049, China 
}

\date{Accepted XXX. Received YYY; in original form ZZZ}

\pubyear{2015}

\begin{document}
\label{firstpage}
\pagerange{\pageref{firstpage}--\pageref{lastpage}}
\maketitle

\begin{abstract}
Galaxy group masses are important to relate these systems with the dark matter halo hosts. However, deriving accurate mass estimates is particularly challenging for low-mass galaxy groups. Moreover, calibration of observational mass-proxies using weak-lensing estimates have been mainly focused on massive clusters. We present here a study of halo masses for a sample of galaxy groups identified according to a spectroscopic catalogue, spanning a wide mass range. The main motivation of our analysis is to assess mass estimates provided by the galaxy group catalogue derived through an abundance matching luminosity technique. We derive total halo mass estimates according to a stacking weak-lensing analysis. Our study allows to test the accuracy of mass estimates based on this technique as a proxy for the halo masses of large group samples. Lensing profiles are computed combining the groups in different bins of abundance matching mass, richness and redshift. Fitted lensing masses correlate with the masses obtained from abundance matching. However, when considering groups in the low- and intermediate-mass ranges, masses computed according to the characteristic group luminosity tend to predict higher values than the determined by the weak-lensing analysis. The agreement improves for the low-mass range if the groups selected have a central early-type galaxy. Presented results validate the use of mass estimates based on abundance matching techniques which provide good proxies to the halo host mass in a wide mass range.
\end{abstract}




\section{Introduction}

Galaxies tend to group together and form galaxy systems 
ranging from galaxy pairs to rich clusters. 
According to the current cosmological $\Lambda$CDM paradigm, 
these systems are expected to reside on highly overdense 
dark matter clumps, called halos. In this context, galaxy systems 
are important to study galaxy evolution 
as well as constrain cosmological parameters within the standard paradigm
\citep[see e.g. ][for reviews]{Allen2011,Kravtsov2012}.
Therefore, reliable and complete group samples spanning a wide range of masses
are important in order to study the evolution of 
these systems and use them as cosmological probes.
 
Commonly adopted galaxy group-finder
algorithms are usually based on photometric 
properties such as photometric redshifts 
\citep[e.g. ][]{Breukelen2009,Milkeraitis2010,Soares-Santos2011,Wen2012,Durret2015,Radovich2017,Bellagamba2018}
or on galaxy detection along the 
red-sequence \citep[e.g. ][]{Gladders2000,Gal2009,Murphy2012,Rykoff2014,Oguri2014,Licitra2016}.
These approaches have the advantage of running on large photometric
data sets providing a large sample of mainly massive ($\gtrsim 5 \times 10^{13}$M$_\odot$)
galaxy groups. On the other hand, identification algorithms based on 
spectroscopic redshift information minimise biases 
introduced by projection effects 
on determining galaxy group memberships. 
Many algorithms based on spectroscopic surveys \citep{Huchra1982,Tucker2000,Merchan2002,Miller2005,Berlind2006,Yang2007,Tempel2012}
have been successfully applied to provide group catalogues 
including systems with a low number of galaxy members
ie. low-mass systems ($\gtrsim 5 \times 10^{12}$M$_\odot$).

Determining the group host halo mass is important in order to use galaxy
systems as cosmological probes and to better characterise them.
Given that the abundance 
and spatial distribution of galaxy systems is connected with
the growth of structures within the cosmic expansion \citep[see e.g][for a review]{Kravtsov2012}, 
comparing the observed distribution of galaxy systems in
halos mass bins to that expected in numerical
simulations, can be used to constrain cosmological parameters.
Moreover, it is expected that the baryonic processes taking place
within the halos are strongly related to their total mass \citep{LeBrun2014}. Hence, 
halo mass estimates are also important to understand the effect of
the environment on galaxy evolution. 

In order to provide suitable group mass estimates, mass-proxies are usually considered including
group richness, and X-ray and optical total luminosity.
These relations are usually calibrated considering the masses
estimated through the application of weak-lensing stacking techniques
\citep[e.g. ][]{Leauthaud2010,Viola2015,Simet2017,Pereira2018,Pereira2020}, since gravitational
lensing provides a direct way to derive the average mass distribution for a sample of galaxy groups.
These stacking techniques are based on the combination of groups within a range of a given observational property such as richness or total luminosity, in order to increase the signal-to-noise
ratio of the lensing signal. 

In general, studies linking weak-lensing halo masses to galaxy systems
have been focused on massive or moderate-massive clusters, since low-mass galaxy groups are 
difficult to identify given their low number of bright members. 
Also, a higher dispersion between a mass-richness relation is expected for low-member 
galaxy groups and a correction to the apparent richness is needed in order to include
faint galaxy members not targeted by the spectroscopic survey.
Moreover, mass estimates are particularly challenging for these systems given that dynamical masses
are not reliable because they are based on a small number of members, and
X-ray luminosity studies are observationally difficult since they
are significantly fainter in comparison to massive systems and, consequently samples are generally small
\citep[e.g.][]{Sun2009,Eckmiller2011,Kettula2013,Finoguenov2015,Pearson2015}. An alternative
approach for mass estimates of low-mass systems comes from the assumption
that there is a one-to-one relation between the characteristic group luminosity 
and the halo mass \citep{Vale2004,Kravtsov2004,Tasitsiomi2004,Conroy2006,Behroozi2010,Cristofari2019}. 
This approach for mass assignment is 
known as the abundance-matching technique and works by ordering the identified systems
according to their characteristic luminosity and associating
masses so that their abundance matches a theoretical mass function. 

In this work we analyse a sample of spectroscopic selected galaxy groups 
identified according to the algorithm presented by \citet{Rodriguez2020}.
The algorithm is based on a combination of percolation and halo-based 
methods. Groups were identified using the spectroscopic data
of the Sloan Digital Sky Survey Data Release 12 \citep[SDSS-DR12,][]{Alam2015}
and spans over a wide range of richness and masses, including a large fraction of
low-richness galaxy systems. Taking into account
the overabundance of low mass systems, the inclusion of these systems in
testing different mass proxies, is important for posterior cosmological analysis that
comprise galaxy systems in a wide mass range.

For our analysis, we select a group sample from this catalogue and performed a weak--lensing analysis in order to estimate mean total halo masses. We consider the brightest
galaxy member (BGM) as the halo centre and we model the possible miscentring effect on the lensing signal considering a fraction
of miscentred groups. We also evaluate the relation between this fraction of groups
and a wrong membership assignment using simulated data. 
Then, we compare derived masses with the estimates
provided in the catalogue, computed according to both the abundance matching technique
and the line-of-sight (LOS) velocity dispersion. We study the relation between
these mass proxies and the lensing estimates and asses to which extent this relation
is biased according to the the group BGM morphology, redshift and richness.
Since the mass proxies provided in the catalogue rely on the membership
assignment by the identification algorithm, the analysis allow us to 
test its performance as well as to study the relation between the total 
halo masses and the mentioned proxies. 

The paper is organised as follows: In Sec. \ref{sec:data} we 
describe the observational and simulated data used in this work,
as well as the galaxy group catalogue. We detail in Sec. \ref{sec:lens}
the weak-lensing stacked analysis performed to derive the total halo masses.
In Sec. \ref{sec:results}, we present the results and study the relation between
the mass proxies provided by the galaxy group catalogue and the lensing
estimates. Finally in Sec. \ref{sec:summary} we summarise and discuss
the results presented in this work. 
When necessary we adopt  a standard cosmological model 
with $H_{0}$\,=\,$70$\,km\,s$^{-1}$\,Mpc$^{-1}$, 
$ \Omega_{m} $\,=\,0.3, and $ \Omega_{\Lambda} $\,=\,0.7.

\section{Data description}
\label{sec:data}

\subsection{Weak-lensing data}

We perform the weak lensing analysis by using a combination of the shear catalogues provided by four public weak-lensing surveys (CFHTLenS, CS82, RCSLenS, and KiDS/KV450)
based on similar quality
observations, which allows the direct combination of these catalogues
 as done in previous works \citep{Gonzalez2020,Xia2020,Schrabback2020}. In this subsection we first briefly describe the lensing surveys in which the shear catalogues are based and the galaxy background selection.
 
 Although the combination of these surveys has been already tested in
previous studies \citep[see Appendix A in ][]{Gonzalez2020}, 
we carried out the lensing analysis using the data of the
individual surveys. Derived lensing masses for the individual surveys
are all in agreement within $2\sigma$ with the combined analysis, 
considering the errors for the individual
mass estimates, and no significant bias are introduced. As complementary material in the Appendix \ref{app:test}, we provide the resulting masses derived from the individual catalogues for the group sample. Masses are binned according to the abundance matching technique.

 \subsubsection{Shear catalogues}
 
 The Canada-France-Hawaii Telescope Lensing Survey (CFHTLenS) weak lensing 
catalogues\footnote{CFHTLenS: http://www.cadc-ccda.hia-iha.nrc-cnrc.gc.
ca/en/community/CFHTLens} are based on observations provided by the CFHT Legacy Survey. This is a multiband survey ($u^*g'r'i'z'$) that spans 154\,deg$^2$ distributed in four separate patches W1, W2, W3 and W4 ($63.8$, $22.6$, $44.2$ and $23.3$ deg$^2$, respectively). Considering a 5$\sigma$ point source detection, the limiting magnitude is $i' \sim 25.5$. The shear catalogue is based on the $i-$band measurements, with a weighted galaxy source density of $\sim 15.1$\,arcmin$^{-2}$. See \citet{Hildebrandt2012,Heymans2012,Miller2013,Erben2013} for further details regarding this shear catalogue. 

The CS82 shear catalogue is based on the observations provided by the CFHT Stripe 82 survey, a joint Canada-France-Brazil project designed
to complement the existing SDSS Stripe 82 $ugriz$
photometry with high-quality $i-$band imaging to be used for lensing measurements \citep{Shan2014,Hand2015,Liu2015,Bundy2017,Leauthaud2017}.
This survey spans over a window of $2 \times 80$\,deg$^2$, with an effective area of $129.2$\,deg$^2$. It has a median point spread function (PSF) of $0.6''$ and a limiting magnitude $i' \sim 24$ \citep{Leauthaud2017}. The source galaxy catalogue has an effective weighted galaxy number density of $\sim 12.3$\,arcmin$^{-2}$ and was constructed using the same weak lensing pipeline developed
by the CFHTLenS collaboration. Photometric
redshifts are obtained using BPZ algorithm from
matched SDSS co-add \citep{Annis2014} and UKIDSS YJHK
\citep{Lawrence2007} photometry.

The RCSLens catalog\footnote{RCSLenS: https://www.cadc-ccda.hia-iha.nrc-cnrc.gc.ca/en/community/rcslens}
\citep{Hildebrandt2016} is based on the Red-sequence Cluster Survey 2
\citep[RCS-2,][]{Gilbank2011}. This is a  multi-band  imaging  survey  in  the $griz-$bands that reaches a depth of $\sim24.3$  in  the $r-$band for a point source at 7$\sigma$ detection level and spans over $\sim785$\,deg$^2$ distributed in 14 patches, the largest being
$10 \times 10$\,deg$^2$ and the smallest $6 \times 6$\,deg$^2$. The source catalogue is based on $r-$band imaging and achieves an effective weighted galaxy number density of $\sim 5.5$\,arcmin$^{-2}$. 

Finally, the KiDS-450 catalog\footnote{KiDS-450:
http://kids.strw.leidenuniv.nl/cosmicshear2018.php}
\citep{Hildebrandt2017} is based on the third
data release of the Kilo Degree
Survey \citep[KiDS,][]{Kuijken2015}, which 
 is a multi-band imaging survey ($ugri$) that spans over 447\,deg$^2$.
Shear catalogues are based on the $r-$band images with a mean
PSF of $0.68''$ and a $5\sigma$ limiting magnitude of $25.0$, resulting in an effective weighted galaxy number density
of $\sim 8.53$\,arcmin$^{-2}$. Shape measurements are performed using an upgraded version of $lens$fit
algorithm \citep{Fenech2017}. 

 These data (except for KiDS-450) are based
on imaging surveys carried-out using the MegaCam 
camera \citep{Boulade2003} mounted on the Canada France Hawaii Telescope (CFHT), therefore, they have similar
image quality. 
In spite that KiDS-450 shear catalogue is based on observations obtained with a different camera, 
both cameras share similar properties, such as a pixel scale of $0.2''$. 
Also, the seeing conditions are similar for all the surveys ($\sim 0.6''$).
Moreover, all the source galaxy catalogues 
were obtained using \textit{lens}fit \citep{Miller2007,Kitching2008}
to compute the shape measurements and photometric redshifts are estimated
using the BPZ algorithm \citep{Benitez2000,Coe2006}.
To combine the catalogues in the overlapping areas we favour (disfavour) CFHTLens (RCSLens) data, since this catalogue is based in the deepest (shallowest) imaging, thus contain the highest (lowest) background galaxy density.

\subsubsection{Galaxy background selection}

For our analysis, we have only included galaxies considering the following \textit{lens}fit parameters cuts: 
MASK $\leq$ 1, FITCLASS $= 0$ and $w > 0$.
Here MASK is a masking flag, FITCLASS is a flag parameter that is set to $0$ when the source is classified as a galaxy and
$w$ is a weight parameter that takes into account errors on the shape measurement and the intrinsic shape noise \citep[see details in ][]{Miller2013}. We carried out the lensing study by applying the additive calibration correction factors for the ellipticity components provided for each catalogue and a multiplicative shear calibration factor to the combined sample of galaxies as suggested by \citet{Miller2013}.

For each group located at a redshift $z$, we select background galaxies, 
i.e. the galaxies that are located behind the group
and thus affected by the lensing effect, taking into account
Z\_BEST$ > z + 0.1$ and ODDS\_BEST $> 0.5$, where
Z\_BEST is the photometric redshift estimated for each galaxy,
and ODDS\_BEST is a parameter that expresses the quality of 
Z\_BEST and takes values from 0 to 1. We also restrict our galaxy background sample by considering the galaxies with Z\_BEST$ > 0.2$ and up to $1.2$ for all the shear catalogues, except for KiDS where a more restrictive cut is taken into account (Z\_BEST$ < 0.9$) according to the suggested by \citet{Hildebrandt2017}. Background galaxies are assigned the  to each group using the public regular grid search algorithm \textsc{grispy}\footnote{https://github.com/mchalela/GriSPy}
\citep{chalela2019grispy}.

\subsection{Galaxy groups}
\label{subsec:groups}

\begin{figure}
    \centering
    \includegraphics[scale=0.55]{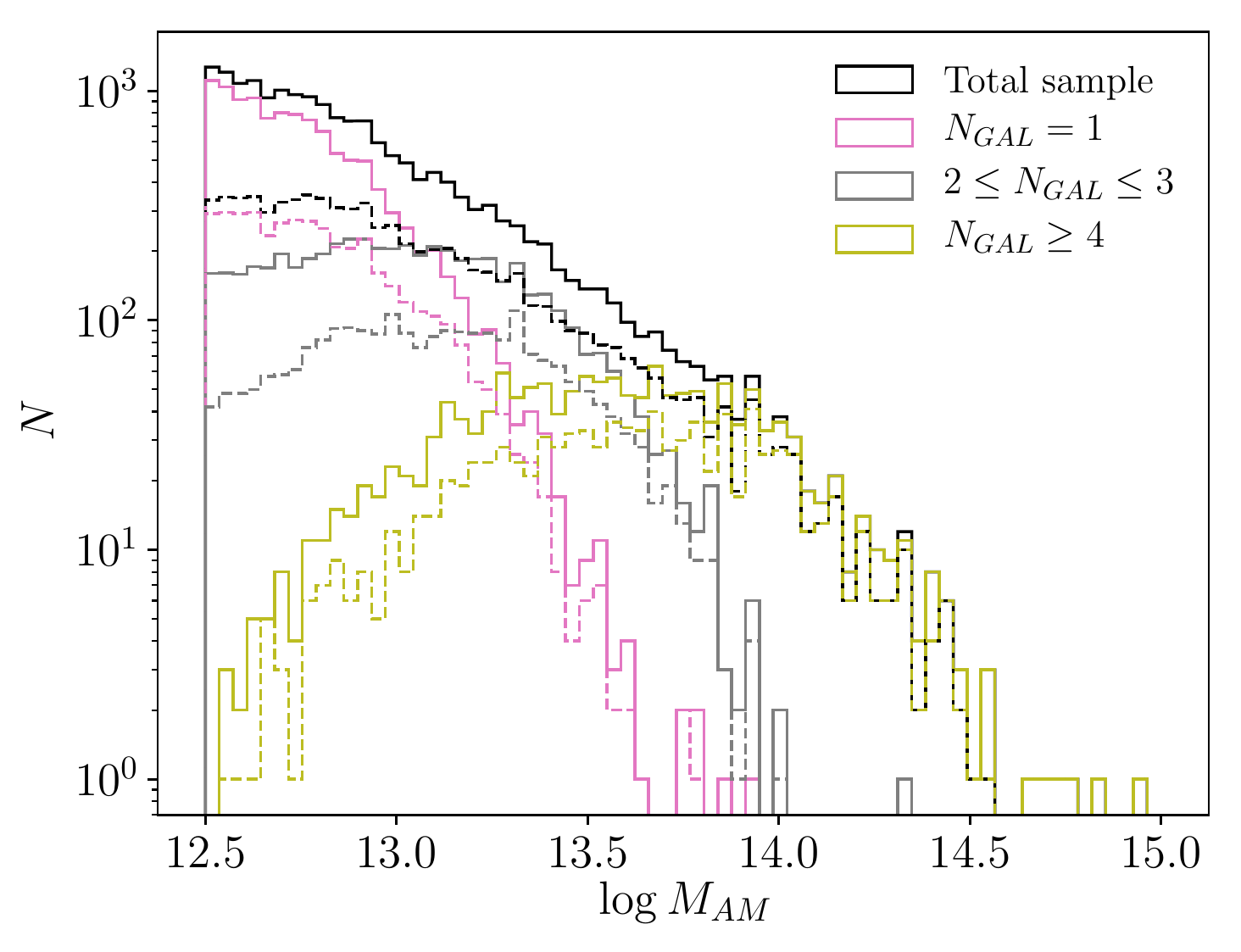}\\
    \includegraphics[scale=0.55]{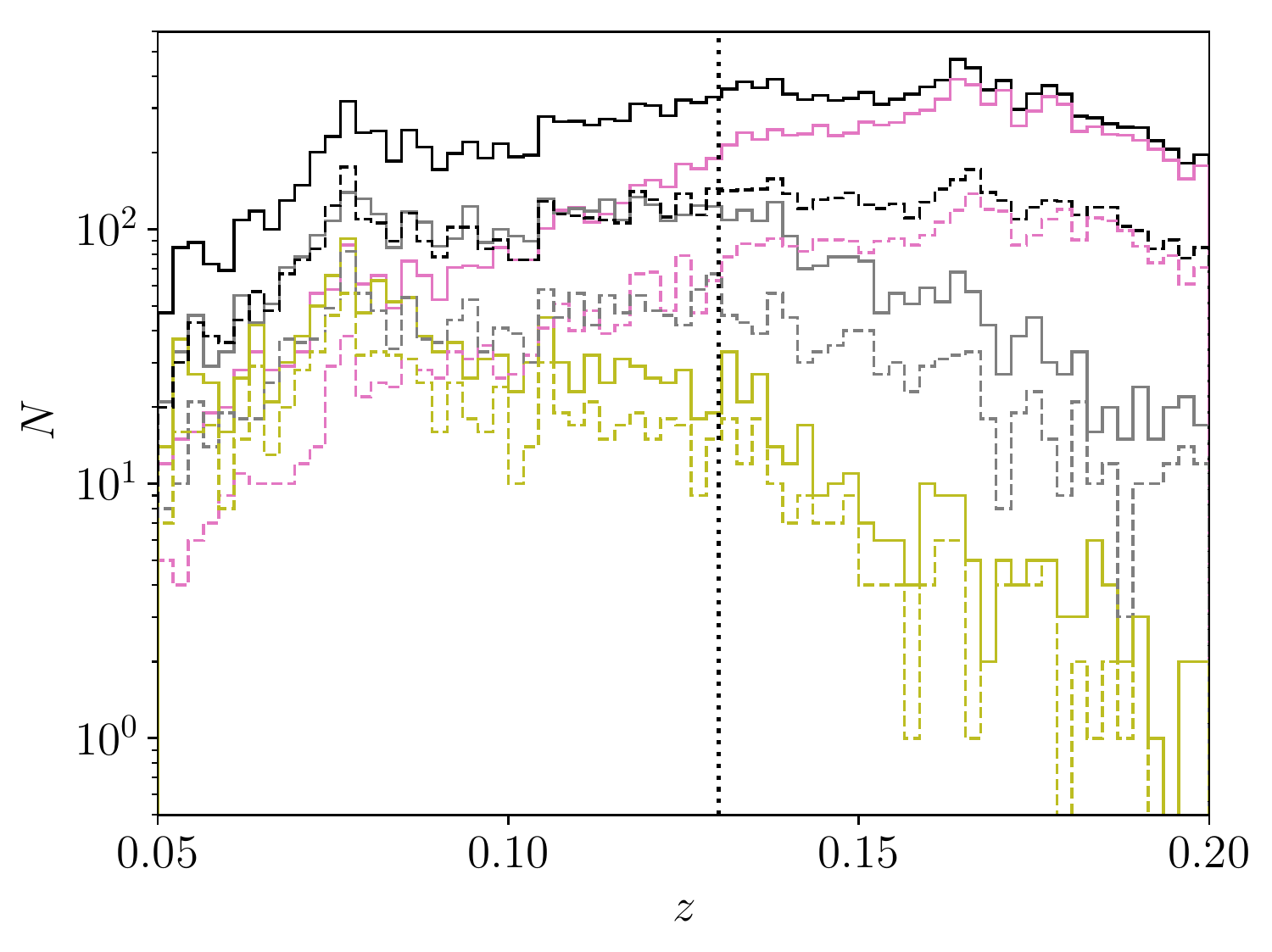}\\
    \includegraphics[scale=0.55]{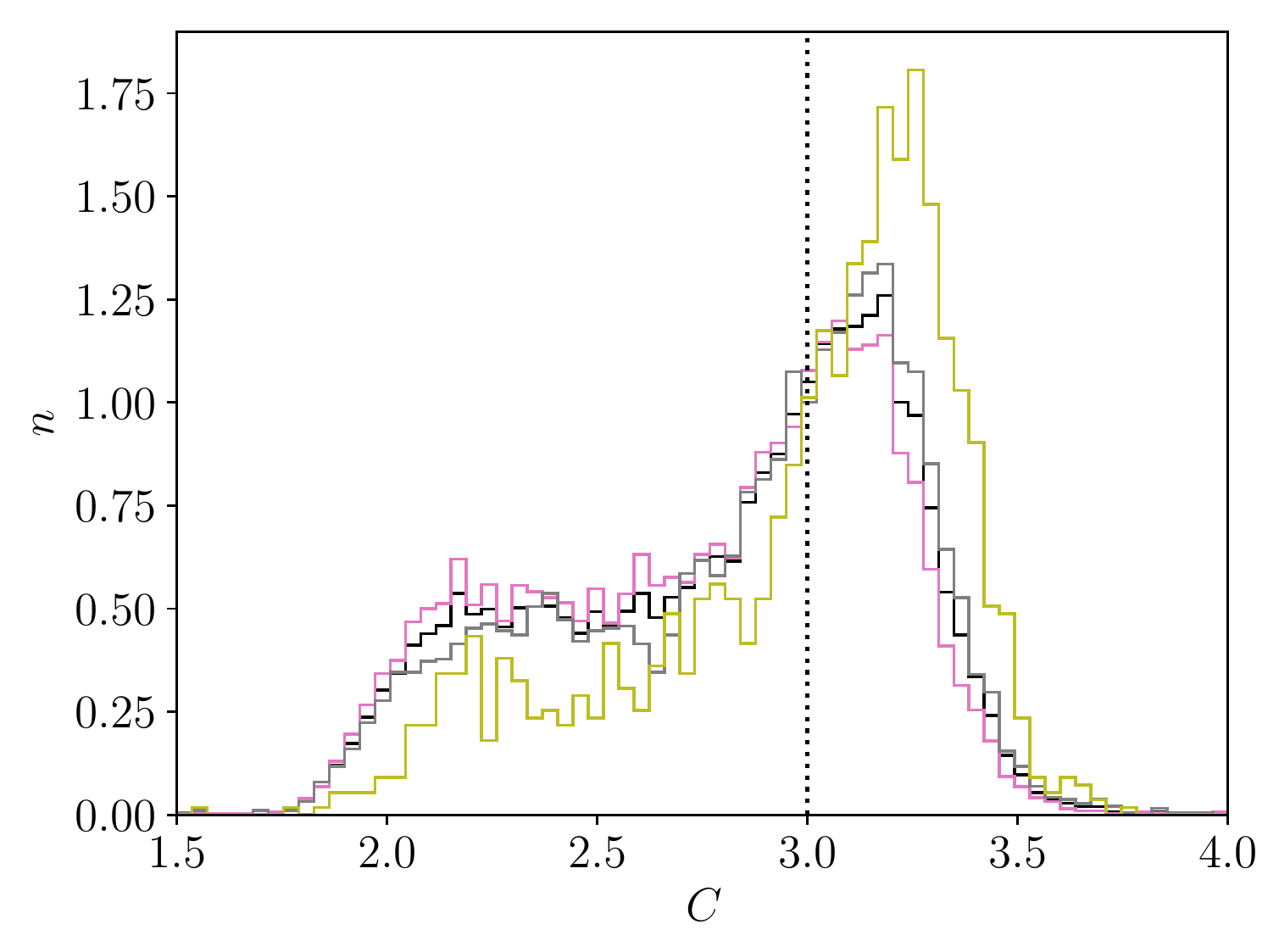}
    \caption{Upper panel and middle panel: Distribution of masses and redshift of
    the analysed samples of groups. Solid and dashed lines correspond to the mass distributions
    obtained for the total and $C-$sample, respectively. Dotted vertical in the middle panel indicates the $z = 0.13$ limit used to select high- and low-redshift samples.
    Lower panel: Normalised distribution of the concentration
    index for the total sample of the groups analysed. The dotted vertical line at $C = 3.0$
    indicates the concentration value adopted to select the $C-$sample of groups. For all the panels
    the coloured distributions correspond to different membership bins selected according to $N_\text{GAL}$.}
    \label{fig:hist}
\end{figure}

We use the publicly available 
galaxy group catalogue\footnote{http:
//iate.oac.uncor.edu/alcance-publico/catalogos/} 
obtained through the identification algorithm by \citet{Rodriguez2020},
 which combines friends-of-friends \citep[FOF, ][]{Huchra1982} and halo-based methods \citep{Yang2005}.
This group finder aims to identify gravitationally bound galaxy systems with at least one bright galaxy, a galaxy with an absolute $r-$band, $M_r$ \footnote{$M_r$ is computed according to the SDSS apparent Petrosian magnitude  and the galaxy spectroscopic redshift, considering the corresponding $k-$correction.}, magnitude lower than $-19.5$. 
By so doing, we consider 
galaxy systems dominated by a central galaxy with fainter members
that were not included in the spectroscopic catalogue.
Briefly, the algorithm performs an iterative identification procedure that consists
in two parts. First, all the galaxies with $M_r< -19.5$ are linked using a FOF method 
based on spatial separation criteria following the
prescriptions in \citet{Merchan2002,Merchan2005}.
After this step, group candidates with at least one bright member are obtained.

Once the catalogue of potential groups is obtained, the membership
assignment is optimised by applying a halo-based group finder following
\citet{Yang2005,Yang2007}. In this step, the algorithm
computes a three-dimensional density contrast in redshift space,
taking into account a characteristic
luminosity calculated according to the potential galaxy members.
The characteristic luminosity, $L_\text{gr}$, associated to each group can be 
estimated from the luminosity of their galaxy members plus a 
correction that takes into account the incompleteness 
due to the limiting magnitude of the observational data \citep{Moore1993}.

Considering each group characteristic luminosity a halo mass is assigned, 
$M_\text{AM}$, performing an abundance matching technique on luminosity
\citep[e.g.,][]{Vale2004,Kravtsov2004,Tasitsiomi2004,Conroy2006}, which assumes a one-to-one relation between the mass and the luminosity. Taking this into account, masses are assigned after
matching the rank orders of the halo masses and their characteristic luminosity
for a given comoving volume, considering the \citet{Warren2006} halo mass function.
A caveat is introduced at this stage, since the assumed halo
mass function is cosmology dependent, which could introduce some biases when
using the density distribution of groups binned in mass as cosmological probes. Nonetheless,
masses can be easily re-computed using another cosmology.
It is important to highlight that, in spite this approach is based in the assumption of a one-to-one relation between the mass and the luminosity, masses are assigned after ranking the groups according to the computed $L_\text{gr}$ and then matching the obtained distribution with the predicted taking into account the halo mass function.

After the mass assignment the algorithm computes the three-dimensional density contrast
assuming that the distribution of galaxies in phase space follows that of the dark matter 
particles and adopting a Navarro–Frenk–White (NFW) profile to compute the projected density.
The density contrast is estimated at the position of each potential
member and only the galaxies that are located above a given threshold 
are considered as belonging to the system. Taking into account 
the new membership assignment, a new characteristic luminosity is computed
and the algorithm iterates until convergence in
the number of members, $N_\text{GAL}$. 

Galaxy groups are obtained by applying the algorithm
to the spectroscopic galaxy catalogue provided by the SDSS-DR12.
The catalogue includes $367370$ groups spanning from 
$z \sim 0.02$ up to $z = 0.3$; of which
$302392$ with one member, $11943$ with
four or more, and a $1386$ with ten or more members. Besides the mass 
estimate, $M_\text{AM}$, assigned during the identification
procedure, the catalogue also
provides for the groups with $N_\text{GAL} \geq 4$ 
the projected LOS velocity dispersion of the group, $\sigma_{V}$ and a dynamical mass estimate, 
$M_{vir}$, computed following \citet{Merchan2002}, 
according to $\sigma_{V}$ and the position of each member.

We restrict the sample to the clusters that are included within the
sky-coverage of the lensing catalogues. We also include in the analysis
only the groups with $\log{M_\text{AM}/(h^{-1} M_\odot)} > 12.5$, to ensure we are considering
group-scale halos, and within a redshift range of $0.05 < z < 0.2$. 
The lower limit in the redshift is selected considering that the lensing signal
decreases for groups at lower redshifts
and the higher limit is selected taking into account that the sample of groups with $N_\text{GAL} \geq 4$
extends up to $z\sim 0.23$.
Applying these criteria, the total sample analysed
comprises of $18030$ systems, $\sim 63 \%$ of these are groups with $N_\text{GAL} = 1$
and $1537$ have more than $4$ members ($N_\text{GAL} \geq 4$). We have considered a
subsample of these groups, hereafter $C-$sample, taking into account the morphology of the 
BGM according to the SDSS concentration index, $C$. This parameter
is usually adopted to separate early- and late-type galaxies samples and
is defined as $C \equiv r_{90}/r_{50}$ (where $r_{90}$ and $r_{50}$ are the radii enclosing $90\%$
and $50\%$ of the $r-$band Petrosian flux, respectively). 
A sample of galaxies with $C > 2.6$ is expected to include $75\%$
of early type galaxies \citep{Strateva2001}. Therefore, we define the $C-$sample
including groups with their BGM having $C \geq 3.0$, which roughly corresponds to the median 
value of the concentration distribution for the total sample of groups analysed. Furthermore, 
in order to explore the effects of variations of scaling relation between mass proxies for group samples with different mean redshift,
we select two samples: a low-redshift ($z < 0.13$), and a high-redshift ($z \geq 0.13$) sample.
In Fig. \ref{fig:hist}
we show the halo mass distribution, $\log{M_\text{AM}}$, for the total sample,
the $C-$sample and other subsamples selected according to the number of members
($N_\text{GAL} = 1$, $2\leq N_\text{GAL} \leq  3$ and $N_\text{GAL} \geq 4$).
We also show the concentration index distribution for the group sample analysed.
As it can be noticed, the concentration cut mainly discard low richness
systems and so, bias the mass distribution to higher values. 

\subsection{Simulated data}
\label{subsec:sim-data}

As will be detailed in Sec. \ref{sec:lens}, for our lensing analysis we assume that the halo centre can be well approximated by the BGM position. In order to test the effects of a wrong membership assignment introduced by the identification algorithm, we use a mock catalogue employing synthetic galaxies extracted from a semi-analytic model of galaxy formation applied on top of the Millennium Run Simulation I \citep{Springel2005}. 

The Millennium Simulation is a cosmological N-body simulation that evolves more than 10 billion dark matter particles in a 500$h^{-1}$Mpc periodic box, using a comoving softening length of 5$h^{-1}$ kpc.  This simulation offers high spatial and time resolution within a large cosmological volume.  This is a dark matter only simulation, but there are different models to populate halos with galaxies. One of these is the semi-analytic galaxy formation model developed by \cite{Guo2010}, that we use to build our synthetic galaxy catalogue.

We construct our catalogue following the same procedure as in \cite{Rodriguez2015} and \cite{Rodriguez2020}. Since the Millenium simulation box is periodic, we place the observer at the coordinate origin and repeat the simulated volume until we reach the SDSS volume. The redshifts were obtained using the distances to the observer and taking into account the distortion produced by proper motions. Finally, to mimic SDSS we impose the same upper apparent magnitude threshold of this catalogue and we use the mask to perform the same angular selection function of the survey.

We obtain the mock galaxy group catalogue by applying \cite{Rodriguez2020} identification algorithm with 
the same criteria as described in the previous subsection. 
In order to compute the shift of the centres due to the identification process, we match the halo to each group identified by our method, by looking for the maximum number of members in common. Then we compute the fraction of groups for which the 
brightest galaxy is located at the halo centre.

\section{Lensing mass estimates}
\label{sec:lens}

\begin{figure*}
    \centering
    \includegraphics[scale=0.65]{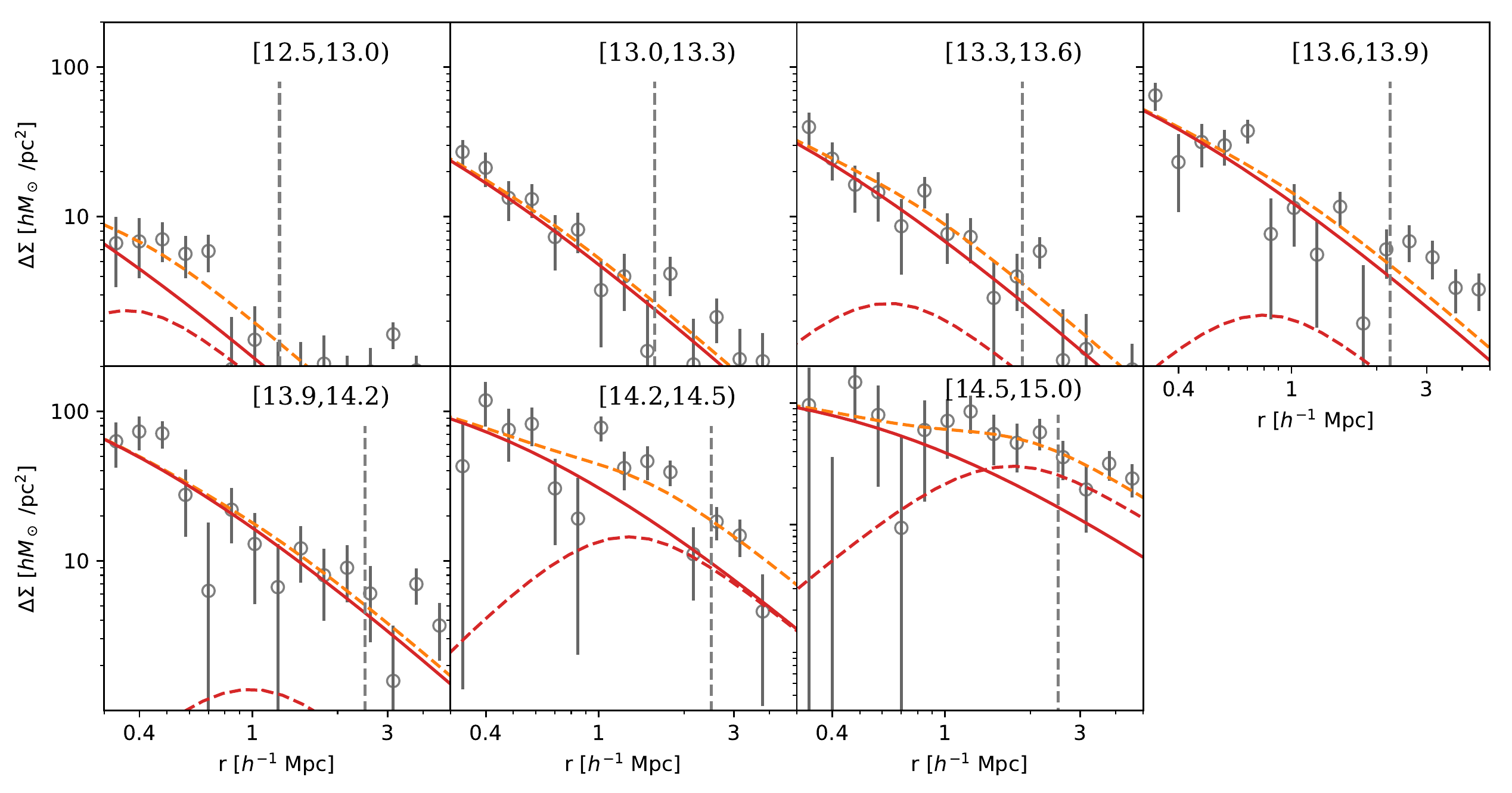}
    \caption{Contrast density profiles for the subsamples selected from the total sample in the whole richness and redshift range ($N_\text{GAL} \geq 1$ and $0.05 \leq z < 0.2$). Labels indicate the $\log{M_\text{AM}}$ bin. Dashed orange lines corresponds to the total fitted profile (Eq. \ref{eq:monomodel}) and solid and dashed red lines corresponds to the centred, $\Delta \Sigma_{cen}$ (Eq. \ref{eq:scen}), and the miscentring terms, $\Delta \Sigma_{mis}$  (Eq. \ref{eq:soff}), respectively. Offset distributions are computed according to Eq. \ref{eq:Pdist}. Vertical dashed lines correspond to the upper limit in the projected radius adopted in the fitting procedure.}
    \label{fig:profiles}
\end{figure*}

\begin{figure*}
    \centering
    \includegraphics[scale=0.65]{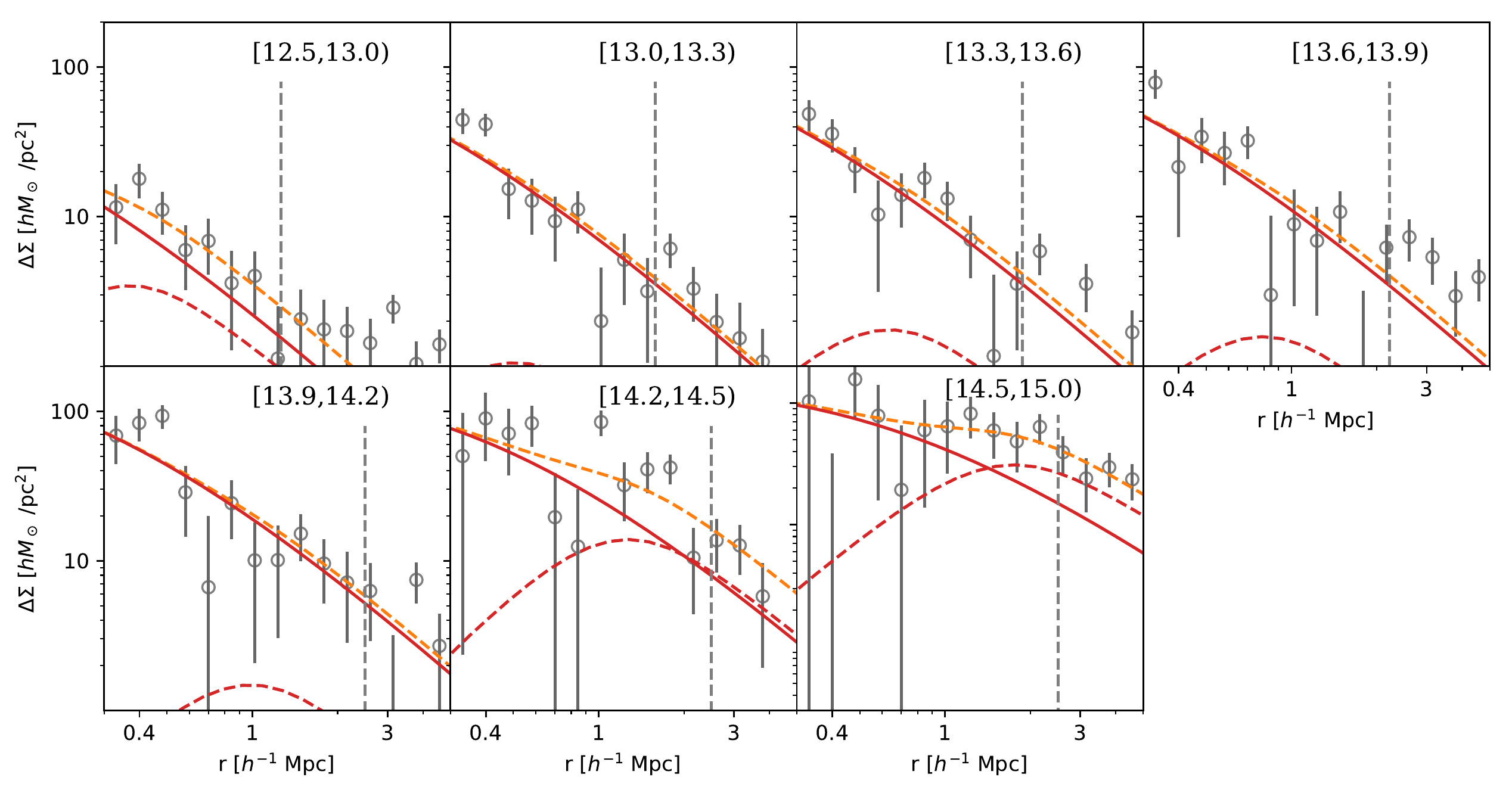}
    \caption{Idem as in Fig. \ref{fig:profiles} but for subsamples selected from the $C-$sample in the whole richness and redshift range ($N_\text{GAL} \geq 1$ and $0.05 \leq z < 0.2$).}
    \label{fig:profiles_cM}
\end{figure*}

\subsection{Adopted formalism}

The weak gravitational lensing effect exerted by the mass distribution associated to galaxy groups, produces a shape distortion of the background galaxies, resulting in an alignment of these galaxies in a tangential orientation with respect to the group centre. The introduced distortion by the lensing effect, can be quantified by the shear parameter, $\gamma = \gamma_1 + i\gamma_2$, and can be estimated according to the measured ellipticity of background galaxies. The observed  ellipticity results in a combination of the galaxy intrinsic shape and the
introduced by the lensing effect. Assuming that the galaxies are randomly orientated in the sky, the shear can be estimated by averaging the ellipticity of many sources, $\langle e \rangle = {\gamma}$.
The noise introduced by the intrinsic shape
of the sources can be reduced by using stacking techniques, which
consists on combining several lenses which 
increase the density of sources. Stacking techniques effectively increase the signal-to-noise ratio of the shear measurements, allowing us to derive reliable average mass density distributions of the combined lenses 
\citep[e.g.][]{Leauthaud2017, Simet2017,Chalela2018, Pereira2020}. 

For a given projected mass density distribution, the azimuthally averaged tangential component, $\gamma_{\rm{t}}$, of the shear can be related with 
the mass density contrast distribution following \citep{Bartelmann1995}:
\begin{equation} \label{eq:DeltaSigma}
    \gamma_{\rm{t}}(r) \times \Sigma_{\rm crit} = \bar{\Sigma}(<r) - \Sigma(r) \equiv \Delta \Sigma(r),
\end{equation}
where we have defined the surface mass density contrast $\Delta\Sigma$. Here $\gamma_{\rm{t}}(r)$ is the tangential component of the shear at a projected distance  $r$ from the centre of the mass distribution, $\Sigma(r)$ is the projected mass surface density distribution, and $\bar{\Sigma}(<r) $ is the average projected mass distribution within a disk at projected distance $r$. $\Sigma_{\rm crit}$ is the critical density defined as:
\begin{equation} \label{eq:sig_crit}
\Sigma_{\rm{crit}} = \dfrac{c^{2}}{4 \pi G} \dfrac{D_{OS}}{D_{OL} D_{LS}},
\end{equation}
where $D_{OL}$, $D_{OS}$ and $D_{LS}$ are  the angular diameter distances from the observer to the lens, from the observer to the source and from the lens to the source, respectively. 

To model the surface density distribution of the halo, $\Sigma$, 
we use the usual NFW profile \citep{Navarro97}. This model depends on two parameters: the radius that encloses the mean density equal to 200 times the critical density of the Universe, 
$R_{200}$, and a dimensionless concentration parameter, $c_{200}$. 
The density profile is defined as:
\begin{equation} \label{eq:nfw}
\rho(r) =  \dfrac{\rho_{\rm crit} \delta_{c}}{(r/r_{s})(1+r/r_{s})^{2}},
\end{equation}
where  $r_{s}$ is the scale radius, $r_{s} = R_{200}/c_{200}$, 
$\rho_{\rm crit}$ is the critical density of the Universe at 
the mean redshift of the sample of stacked galaxy groups, $\langle z \rangle$, 
and $\delta_{c}$ is the cha\-rac\-te\-ris\-tic overdensity of the halo:
\begin{equation}
\delta_{c} = \frac{200}{3} \dfrac{c_{200}^{3}}{\ln(1+c_{200})-c_{200}/(1+c_{200})}.  
\end{equation}
We compute $\langle z \rangle$ by averaging the group sample redshifts, weighted according to the number of background galaxies considered for each group.
The mass within $R_{200}$ can be obtained as 
\mbox{$M_{200}=200\,\rho_{\rm crit} (4/3) \pi\,R_{200}^{3}$}. 
The lensing formula adopted to model this profile is described by \citet{Wright2000}. 
In the fitting procedure we use a fixed mass-concentration relation 
$c_{200}(M_{200},z)$, derived from simulations by \citet{Duffy2008}: 
\begin{equation}
c_{200}=5.71\left(M_{200}/2 \times 10^{12} h^{-1}\right)^{-0.084}(1+\langle z \rangle)^{-0.47}.
\end{equation}
This approach is applied since the concentration parameter mainly affects the slope in the inner regions of the profile and therefore is poorly constrained. Nevertheless, as shown in previous studies the particular choice of this relation does not have a significant impact on the final mass values, which have uncertainties dominated by the noise of the shear profile \citep{Rodriguez2020b}.

To compute the profiles we adopt as the group centre, the position of the BGM. 
The offset distribution between these galaxy-based centres and the true halo centre
can be described by considering two group sample populations: well-centred
and miscentred groups \citep{Yang2006,Johnston2007,Ford2014,Yan2020}.
This miscentring affects the observed shear profile by flattening the lensing signal at the inner regions.
We consider the miscentring effect in our analysis by modelling the contrast density distribution
taking into account two terms: 
\begin{equation} \label{eq:monomodel}
    \Delta \Sigma = p_\text{cc}\Delta \Sigma_{cen} + (1-p_\text{cc}) \Delta \Sigma_{mis},
\end{equation}
where $\Delta \Sigma_{cen}$ and $\Delta \Sigma_{mis}$ 
correspond to the contrast density distribution for 
a perfectly centred, and a miscentred dark matter distribution, respectively, and 
$p_\text{cc}$ is the fraction of well-centred clusters. $\Delta \Sigma_{cen}$ is obtained as:
\begin{equation} \label{eq:scen}
  \Delta \Sigma_{cen}(r) = \bar{\Sigma}(<r) - \Sigma(r). 
\end{equation}

The miscentring term, $\Delta \Sigma_{mis}$, 
is modelled following \citet{Yang2006,Johnston2007,Ford2014}. 
An axis-symmetric surface mass density distribution
whose centre is offset by $r_{s}$, with respect to the adopted centre in the lens plane, results in a projected average density profile given by:
\begin{equation}
    \Sigma(r|r_s) = \frac{1}{2\pi} \int^{2\pi}_0 \Sigma \left( \sqrt{r^2 + r^2_s + 2 r r_s \cos{\theta}} \right) d\theta.
\end{equation}
The fraction of miscentered groups is expected to be shifted following a Gaussian distribution, therefore
the projected offsets can be modelled according to a Rayleigh distribution:
\begin{equation}
\label{eq:Pdist}
    P(r_s) = \frac{r_s}{\sigma^2_\text{off}} \exp{\left(-\frac{1}{2}\frac{r^2_{s}}{\sigma^2_\text{off}}\right)}.
\end{equation}
Alternatively, we have also considered a Gamma function with shape parameter $k=2$ to model the offset distribution:
\begin{equation}
\label{eq:Pdistg}
    P(r_s) = \frac{r_s}{\sigma^2_\text{off}} \exp{\left(-\frac{r_{s}}{\sigma_\text{off}}\right)}.
\end{equation}
We consider this model based on the recent work of \citet{Yan2020}
where they study the miscentring effect by considering different proxies using hydrodynamic simulations.

Taking this into account, the miscentred density can be computed as follows:
\begin{equation} \label{eq:off}
    \Sigma_{mis}(r) = \int_{0}^\infty P(r_s) \Sigma(r|r_s) dr,
\end{equation}
such that the miscentring term for the density contrast profile is:
\begin{equation} \label{eq:soff}
    \Delta \Sigma_{mis}(r) = \bar{\Sigma}_{miss}(<r) - \Sigma_{mis}(r).
\end{equation}

\subsection{Computed estimator and fitting procedure}

We compute the density contrast distribution profiles by averaging
the tangential ellipticity component of the background galaxies of each group
considered in the stacking, as:
\begin{equation}
 \Delta \tilde{\Sigma}(r) = \frac{\sum_{j=1}^{N_L} \sum_{i=1}^{N_{S,j}} \omega_{LS,ij} \Sigma_{{\rm crit},ij} e_{{\rm t},ij}}{\sum_{j=1}^{N_L} \sum_{i=1}^{N_{S,j}} \omega_{LS,ij}},
\end{equation}
where $\omega_{LS,ij}$ is the inverse variance weight computed according to the weight, $\omega_{ij}$, given
by the $lens$fit algorithm for each background galaxy, $\omega_{LS,ij}=\omega_{ij}/\Sigma^2_{{\rm crit},ij}$. 
$N_L$ is the number of galaxy groups considered for the stacking and 
$N_{S,j}$ the number of background galaxies located at a distance 
$r \pm \delta r$ from the $j$th group. 
$\Sigma_{{\rm crit},ij}$ is the critical density for the $i-$th background galaxy of the $j-$th group. The inner regions of the profile could be affected by a stellar mass contribution of the central galaxies. Moreover, these 
regions are more affected by the background selection and an increased scatter to low sky area in the inner regions. Taking these facts into account, we obtain the profiles by binning the background galaxies in 15 non-overlapping log-spaced $r$ bins, from $300 h^{-1}$kpc up to $5 h^{-1}$Mpc.

Errors in the photometric redshifts can led to the inclusion of foreground or galaxy members in the background galaxy sample. These galaxies are unlensed and result in an underestimated density contrast, which is called as the dilution effect. In order to take this effect into account, the $\Delta \tilde{\Sigma}$ measurement can be boosted to recover the corrected signal by using the so-called \textit{boost-factor} \citep{Kneib2003,Sheldon2004,Applegate2014,Hoekstra2015,Simet2017,Leauthaud2017,Melchior2017,McClintock2019,Varga2019,Pereira2020}: $1/(1-f_{cl})$, where $f_{cl}$ is the cluster contamination fraction and it is expected to be higher in the inner radial bins where the contamination by cluster members is more significant. We compute $f_{cl}$ by using a similar approach as the presented in \citet{Hoekstra2007}. Since a non-contaminated background galaxy sample will present a constant density for all the considered radial bins, by computing the excess in the density at each considered radial bin we obtain an estimated value of $f_{cl}$. This excess is computed taking into account the background galaxy density obtained for the last radial bin at  $5 h^{-1}$Mpc, where the contamination of unlensed galaxies is expected to be negligible. By doing so, we obtain the $f_{cl}(r)$ fraction which is included in the analysis. The inclusion of the \textit{boost-factor} in the analysis result in higher mass estimates by a $\sim10\%$ ($\sim25\%$) for the lowest (highest) mass bin sample of groups.

In order to estimate the group halo masses
we fit the computed profiles with the adopted model (Eq. \ref{eq:monomodel})
considering the two free parameters, $p_\text{cc}$ and $M_\text{WL}$ (where $M_\text{WL}$ 
is the $M_{200c}$ mass). We fix the width of the offset distributions, $\sigma_\text{off}$, in terms of the radius, $R_{AM}$, which is the $R_{200}$ radius estimated from the abundance matching mass $M_\text{AM}$, which are expected to be related \citep{Simet2017}. Thus, we set $\sigma_\text{off} = 0.4 \times R_{AM}$
in Eq. \ref{eq:Pdist} according to the results presented in \citet{Simet2017}.
On the other hand, the offset dispersion in  Eq. \ref{eq:Pdistg}
is given by $\sigma_\text{off} = 0.3 \times R_{AM}$ from \citet{Yan2020}.
This approach is similar to the one applied in the fitting procedure of previous stacking analysis \citep{Simet2017,Pereira2018,McClintock2019,Pereira2020} in which
$\sigma_\text{off}$ is fitted considering a radius computed according to the
richness estimator. Although this parameter is fixed taking into account
the radius derived according to the $M_\text{AM}$ mass estimate, we also try
fitting this parameter together with the mass estimate, $\sigma_\text{off} = 0.4 \times R_{WL}$ and $\sigma_\text{off} = 0.3 \times R_{WL}$. The resultant fitted parameters, $p_\text{cc}$ and $M_\text{WL}$, were in agreement with the previous estimates but less constrained. Therefore our final masses do not strongly depend on $\sigma_\text{off}$. We highlight that the fitted miscentring term can be also affected by the adopted concentration, $c_{200}$, since this parameter impacts in the slope profile. We also neglect the contribution of the 2-halo term, introduced by the contribution of neighbouring halos, by fitting the profiles up to a limiting projected radius of $R_{OUT}$. This radius is estimated according to the relation presented by \citet{Simet2017} to compute the upper limit radius taking into account $M_\text{AM}$. For the highest mass bins considered in the analysis, where the lack of modelling of the 2-halo term can biases the lensing mass estimates, we adopt a more restrictive limiting radius of $2.5h^{-1}$Mpc. Thus, all the profiles are fitted up to $\min(R_{OUT},2.5h^{-1}$Mpc).

We constrain our free parameters, $p_\text{cc}$ and $M_\text{WL}$, by using the Markov chain Monte Carlo (MCMC) method, implemented through \texttt{emcee} python package \citep{Foreman2013} to optimise the  log-likelihood 
function for the density contrast profile, $\ln{\mathcal{L}}(\Delta \Sigma | r ,M_\text{WL},p_\text{cc})$.  We fit the data by using 10 walkers for each parameter and 500 steps, considering flat priors for the mass and the fraction of well-centred groups, $11.5 < \log(M_\text{WL}/(h^{-1} M_\odot)) < 16$ and $0.2 < p_\text{cc} < 1.0$. We adopt as the best fit parameters  the median value of the posterior distributions and the correspondent errors are based on the differences between the median and the $16^{th}$ and $84^{th}$ percentiles, without considering the first 100 steps of each chain. We show in Fig. \ref{fig:profiles} and \ref{fig:profiles_cM} the computed profiles together with the fitted models, for the subsamples selected in $\log M_\text{AM}$ bins from the total sample and the $C-$sample, respectively.

\section{Results}
\label{sec:results}

In this section, we first discuss the adopted miscentring modelling
and compare the lensing results to numerical simulations. 
Then, we compare the derived lensing masses to the mass 
estimates provided by the group catalogue, $M_\text{AM}$, computed
according to the abundance matching assignment.
We also study biases in group masses for the different subsamples 
selected considering the group richness and redshift. 
Finally, for the groups with $N_\text{GAL} \geq 4$,
we compare derived lensing mass estimates to the projected LOS velocity dispersion, $\sigma_V$.

In order to study the relation between the abundance matching masses
and the lensing estimates, we split the total sample and the $C-$sample 
of groups in seven $\log{M_\text{AM}}$ bins
from $10^{12.5} h^{-1} M_\odot $ up to $10^{15} h^{-1} M_\odot $. We also obtain
the lensing masses considering the richness subsamples defined in 
subsection \ref{subsec:groups} and high- and low-redshift subsamples. 
In Table \ref{table:Nresults} and \ref{table:results} we describe the selection criteria together with 
the best fitted parameters for the samples selected according to the richness and redshift, respectively. In Appendix \ref{app:corner} we show the 2D posterior probability distributions for the total sample. We also show in Appendix \ref{app:lum}
the characteristic luminosity distributions, their medians and 15- and 85-th percentiles for each considered bin. Derived lensing masses for the sub-samples range from $3 \times 10^{12} h^{-1} M_\odot$ to $5 \times 10^{14} h^{-1} M_\odot$. 
Therefore, our analysis spans over a wide range of halo masses. We further discuss the results obtained in the next subsections.
\begin{table*}
\begingroup
\setlength{\tabcolsep}{10pt} 
\renewcommand{\arraystretch}{1.5} 
\centering
\caption{Fitted parameters for the analysed galaxy groups in the whole redshift range ($0.05 \leq z < 0.2$).}
\begin{tabular}{c c | c c c | c c c}
\hline
\hline
\rule{0pt}{1.05em}%
Richness  & $\log{M_\text{AM}}$ & \multicolumn{3}{c|}{Total sample} & \multicolumn{3}{c}{$C-$sample} \\
selection       &                                   & $N_L $    &  $M_\text{WL}$ & $p_\text{cc}$ & $N_L $    &  $M_\text{WL}$ & $p_\text{cc}$  \\
                  &   [$\log{(h^{-1} M_\odot)}$]    &       &   [$10^{13} h^{-1} M_\odot$] & & & [$10^{13} h^{-1} M_\odot$]          &   \\
\hline
\rule{0pt}{1.05em}   
$N_\text{GAL} \geq 1$ & $[12.5,13.0)$ & $12647$ & $0.31_{-0.07}^{+0.05}$ & $0.57_{-0.25}^{+0.26}$ & $4421$ & $0.60_{-0.14}^{+0.10}$ & $0.62_{-0.28}^{+0.26}$ \\ 
& $[13.0,13.3)$ & $3093$ & $0.95_{-0.15}^{+0.13}$ & $0.90_{-0.14}^{+0.08}$ & $1551$ & $1.48_{-0.22}^{+0.19}$ & $0.91_{-0.15}^{+0.07}$ \\ 
& $[13.3,13.6)$ & $1406$ & $1.81_{-0.28}^{+0.27}$ & $0.75_{-0.18}^{+0.15}$ & $832$ & $2.14_{-0.39}^{+0.30}$ & $0.85_{-0.17}^{+0.11}$ \\ 
& $[13.6,13.9)$ & $571$ & $3.4_{-0.5}^{+0.5}$ & $0.83_{-0.15}^{+0.12}$ & $380$ & $2.80_{-0.63}^{+0.51}$ & $0.86_{-0.15}^{+0.10}$ \\ 
& $[13.9,14.2)$ & $236$ & $4.5_{-0.8}^{+0.9}$ & $0.88_{-0.16}^{+0.09}$ & $178$ & $5.3_{-1.0}^{+1.0}$ & $0.89_{-0.16}^{+0.08}$ \\ 
& $[14.2,14.5)$ & $68$ & $21_{-4}^{+4}$ & $0.50_{-0.14}^{+0.17}$ & $50$ & $18_{-4}^{+5}$ & $0.47_{-0.15}^{+0.16}$ \\ 
& $[14.5,15.0)$ & $9$ & $61_{-25}^{+19}$ & $0.32_{-0.08}^{+0.14}$ & $7$ & $66_{-23}^{+19}$ & $0.32_{-0.09}^{+0.17}$ \\ 
\hline
  $N_\text{GAL} =1$            & $[12.5,12.9)$ & $8815$ & $0.34_{-0.09}^{+0.07}$ & $0.52_{-0.28}^{+0.25}$ &$2894$ & $0.56_{-0.15}^{+0.14}$ & $0.65_{-0.28}^{+0.25}$ \\ 
& $[12.9,13.1)$ & $1703$ & $0.70_{-0.20}^{+0.17}$ & $0.75_{-0.28}^{+0.16}$ &$810$ & $1.35_{-0.39}^{+0.25}$ & $0.79_{-0.28}^{+0.16}$ \\ 
& $[13.1,13.5)$ & $748$ & $1.70_{-0.40}^{+0.37}$ & $0.82_{-0.20}^{+0.12}$ &$442$ & $2.17_{-0.51}^{+0.45}$ & $0.84_{-0.20}^{+0.12}$ \\ 
\hline                           
$2 \leq N_\text{GAL} \leq 3$  & $[12.5,13.5)$ & $4843$ & $0.56_{-0.11}^{+0.12}$ & $0.57_{-0.25}^{+0.18}$ &$2069$ & $1.12_{-0.22}^{+0.20}$ & $0.73_{-0.25}^{+0.18}$ \\ 
& $[13.5,14.5)$ & $357$ & $3.8_{-0.6}^{+0.7}$ & $0.79_{-0.18}^{+0.07}$ &$237$ & $3.7_{-0.7}^{+0.7}$ & $0.90_{-0.18}^{+0.07}$ \\ 
\hline                           
  $N_\text{GAL} \geq 4$      & $[12.5,13.8)$ & $1113$ & $1.29_{-0.31}^{+0.28}$ & $0.68_{-0.22}^{+0.14}$ &$639$ & $1.31_{-0.36}^{+0.33}$ & $0.81_{-0.22}^{+0.14}$ \\ 
& $[13.8,14.2)$ & $348$ & $3.9_{-0.6}^{+0.7}$ & $0.90_{-0.14}^{+0.07}$ &$255$ & $4.8_{-0.8}^{+0.7}$ & $0.90_{-0.14}^{+0.07}$ \\ 
& $[14.2,15.5)$ & $75$ & $28_{-5}^{+4}$ & $0.42_{-0.10}^{+0.17}$ &$56$ & $26_{-5}^{+4}$ & $0.36_{-0.10}^{+0.17}$ \\ 
\hline         
\end{tabular}
\medskip
\begin{flushleft}
\textbf{Notes.} Columns: (1) Richness range of the selected sub-samples (2) Selection criteria according to the abundance matching mass, $M_\text{AM}$; (3), (4) and (5)  number of groups considered in the stacked sample and fitted parameters, $M_\text{WL}$ and $p_\text{cc}$, for the total sample of groups. (6), (7) and (8) same for the groups included in the $C-$sample.
\end{flushleft}
\label{table:Nresults}
\endgroup
\end{table*}

\begin{table*}
\begingroup
\setlength{\tabcolsep}{10pt} 
\renewcommand{\arraystretch}{1.5} 
\centering
\caption{Fitted parameters for the analysed galaxy groups in the whole richness range ($N_\text{GAL} \geq 1$).}
\begin{tabular}{c c | c c c | c c c}
\hline
\hline
\rule{0pt}{1.05em}%
Redshift  & $\log{M_\text{AM}}$ & \multicolumn{3}{c|}{Total sample} & \multicolumn{3}{c}{$C-$sample} \\
selection       &                                   & $N_L $    &  $M_\text{WL}$ & $p_\text{cc}$ & $N_L $    &  $M_\text{WL}$ & $p_\text{cc}$  \\
                  &   [$\log{(h^{-1} M_\odot)}$]    &       &   [$10^{13} h^{-1} M_\odot$] & & & [$10^{13} h^{-1} M_\odot$]          &   \\
\hline
\rule{0pt}{1.05em}   
$z < 0.13$ & $[12.5,13.0)$ & $5186$ & $0.25_{-0.12}^{+0.08}$ & $0.56_{-0.25}^{+0.28}$ & $1963$ & $0.41_{-0.22}^{+0.14}$ & $0.57_{-0.26}^{+0.30}$ \\ 
& $[13.0,13.3)$ & $1317$ & $0.70_{-0.22}^{+0.18}$ & $0.69_{-0.27}^{+0.22}$ & $679$ & $1.28_{-0.35}^{+0.28}$ & $0.85_{-0.20}^{+0.11}$ \\
& $[13.3,13.6)$ & $639$ & $1.18_{-0.39}^{+0.30}$ & $0.74_{-0.27}^{+0.18}$ & $389$ & $1.34_{-0.50}^{+0.38}$ & $0.72_{-0.25}^{+0.20}$ \\ 
& $[13.6,13.9)$ & $288$ & $1.58_{-0.68}^{+0.54}$ & $0.65_{-0.28}^{+0.23}$ & $206$ & $1.54_{-0.75}^{+0.61}$ & $0.63_{-0.27}^{+0.23}$ \\ 
& $[13.9,14.2)$ & $144$ & $4.1_{-1.0}^{+0.9}$ & $0.86_{-0.17}^{+0.10}$ & $112$ & $4.5_{-1.2}^{+1.1}$ & $0.89_{-0.15}^{+0.08}$ \\ 
& $[14.2,14.5)$ & $43$ & $22_{-6}^{+5}$ & $0.42_{-0.13}^{+0.19}$ & $32$ & $20_{-6}^{+5}$ & $0.42_{-0.15}^{+0.23}$ \\ 
& $[14.5,15.0)$ & $6$ & $14_{-54}^{+15}$ & $0.57_{-0.27}^{+0.28}$ & $4$ & $10.0_{-45.9}^{+15.7}$ & $0.60_{-0.26}^{+0.26}$ \\ 
\hline
$z \geq 0.13$ & $[12.5,13.0)$ & $7461$ & $0.31_{-0.08}^{+0.07}$ & $0.69_{-0.30}^{+0.21}$ &$2458$ & $0.64_{-0.15}^{+0.12}$ & $0.72_{-0.30}^{+0.21}$ \\ 
& $[13.0,13.3)$ & $1776$ & $1.26_{-0.21}^{+0.18}$ & $0.86_{-0.20}^{+0.11}$ &$872$ & $1.80_{-0.41}^{+0.29}$ & $0.84_{-0.22}^{+0.12}$ \\ 
& $[13.3,13.6)$ & $767$ & $2.13_{-0.43}^{+0.43}$ & $0.70_{-0.24}^{+0.20}$ &$443$ & $2.83_{-0.52}^{+0.51}$ & $0.79_{-0.19}^{+0.15}$ \\ 
& $[13.6,13.9)$ & $283$ & $5.3_{-0.8}^{+0.7}$ & $0.90_{-0.14}^{+0.08}$ &$174$ & $4.8_{-0.8}^{+0.9}$ & $0.88_{-0.15}^{+0.09}$ \\ 
& $[13.9,14.2)$ & $92$ & $5.6_{-1.8}^{+1.6}$ & $0.72_{-0.24}^{+0.19}$ &$66$ & $7.2_{-2.2}^{+2.0}$ & $0.74_{-0.23}^{+0.18}$ \\ 
& $[14.2,14.5)$ & $25$ & $21_{-6}^{+6}$ & $0.67_{-0.20}^{+0.21}$ &$18$ & $20_{-7}^{+6}$ & $0.67_{-0.25}^{+0.22}$ \\ 
& $[14.5,15.0)$ & $3$ & $68_{-30}^{+23}$ & $0.37_{-0.11}^{+0.20}$ &$3$ & $66_{-29}^{+22}$ & $0.39_{-0.13}^{+0.25}$ \\ 
\hline         
\end{tabular}
\medskip
\begin{flushleft}
\textbf{Notes.} Columns: (1) Redshift range of the seleceted subsamples (2) Selection criteria according to the abundance matching mass, $M_\text{AM}$; (3), (4) and (5)  number of groups considered in the stacked sample and fitted parameters, $M_\text{WL}$ and $p_\text{cc}$, for the total sample of groups. (6), (7) and (8) same for the groups included in the $C-$sample.
\end{flushleft}
\label{table:results}
\endgroup
\end{table*}

\subsection{Miscentring study}

\begin{figure}
    \centering
    \includegraphics[scale=0.55]{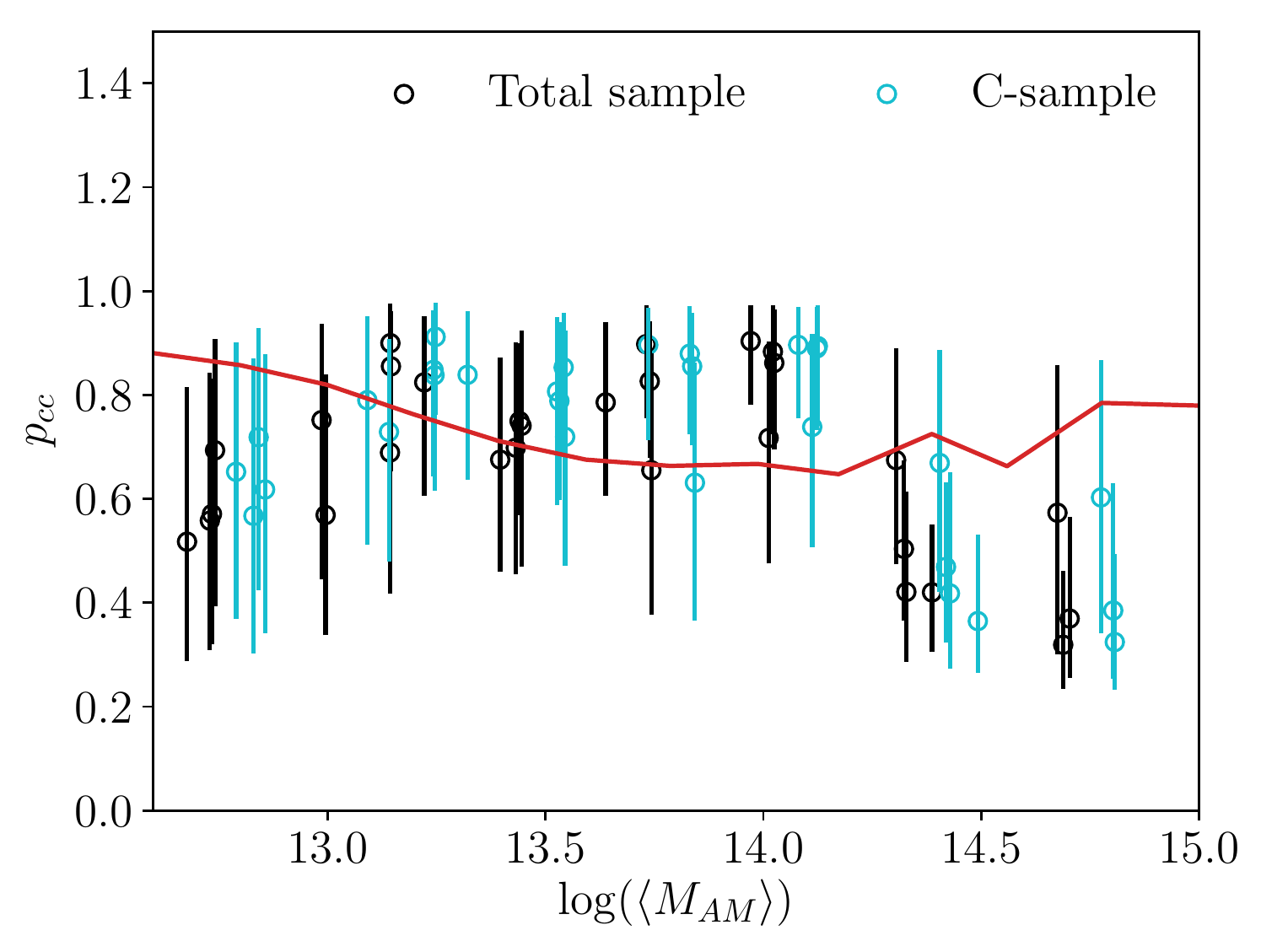}
    \caption{Fraction of well-centred galaxy groups, $p_\text{cc}$, fitted according to the 
    density contrast profiles for the considered samples detailed in Table \ref{table:Nresults}
    and Table \ref{table:results}
    as a function of the mean $M_\text{AM}$. The red solid line is the fraction obtained from the
    simulated data, that considers the expected miscentered introduced by a wrong membership
    assignment. For the $C-$samples values are shifted in the x-axis by $0.1$ for a better visualisation of the Figure.}
    \label{fig:pcc}
\end{figure}

Taking into account the two different expressions (Eq. \ref{eq:Pdist}
and \ref{eq:Pdistg}) to model the offset distribution of the miscentred
groups, we fit two sets of free parameters, $p_\text{cc}$ and $M_\text{WL}$,
for each model. We find no significant differences between the 
reduced chi-square values obtained from both offset modelling, obtaining a mean
of the reduced chi-square ratios of $0.99$, and a standard deviation of $0.02$. 
Moreover, the fitting parameters are in excellent mutual agreement, since the 
mean ratio of $M_\text{WL}$ ($p_\text{cc}$)  is $1.05$ ($1.04$)
with a standard deviation of $0.05$ ($0.08$). Therefore, both 
modellings provide consistent profiles within the fitting parameter uncertainties. 
For the rest of our analysis we consider only the parameters derived
taking into account Eq. \ref{eq:Pdist}. 

In order to test
if the results are consistent with the expected miscentring due
to a wrong membership assignation by the identification algorithm, 
we use the mock sample of groups described in subsection \ref{subsec:sim-data}.
We compute the projected distance distribution between the central galaxy of the halo 
and the brightest member assigned to the group hosted by the halo.
Then, we compute the fraction of well-centred groups in bins of $\log{M_\text{AM}}$,
i.e. the fraction of groups of which the BGM is the central galaxy of the 
dark matter host halo. We also fit Equations \ref{eq:Pdist}
and \ref{eq:Pdistg} to the distributions of projected distances, to estimate the dispersion, $\sigma_\text{off}$.
Estimated dispersion values of the Rayleigh distribution are systematically
higher than the dispersion fitted using a Gamma distribution by a factor $\sim1.4$,
with a mean $\sigma_\text{off}$ of 0.4 and 0.3, respectively. This supports the adopted
fixed values for $\sigma_\text{off}$ stated in the previous section. 

In Fig. \ref{fig:pcc} we show the $p_\text{cc}$ values obtained from the lensing analysis 
together with that derived from the mock sample as a function of the mean
$M_\text{AM}$ for each bin. No significant differences are obtained when considering
the $C-$sample. For the observed group sample, as well as for
the groups identified in the simulation, the fraction of well-centred groups
tend to decrease with the mean mass. Although there is a general agreement
between $p_\text{cc}$ estimates derived from the simulated sample and the lensing estimates,
these later estimates tend to be systematically biased to lower values for massive systems ($>  10^{14} h^{-1} M_\odot$) which are expected to include a larger
fraction of merging systems. It is important to highlight that the analyses based on the simulated data
only takes into account the miscentred introduced by errors in the membership
assignation, thus, it does not considers possible offsets
between the dark matter halo and the BGM centres due to gas/galaxy dynamics. On the other
hand, lensing $p_\text{cc}$ values result from a combination of both effects. 

\subsection{Relating lensing masses to the abundance matching prediction}
\label{subsec:mrelation}

\begin{figure*}
    \centering
    \includegraphics[scale=0.55]{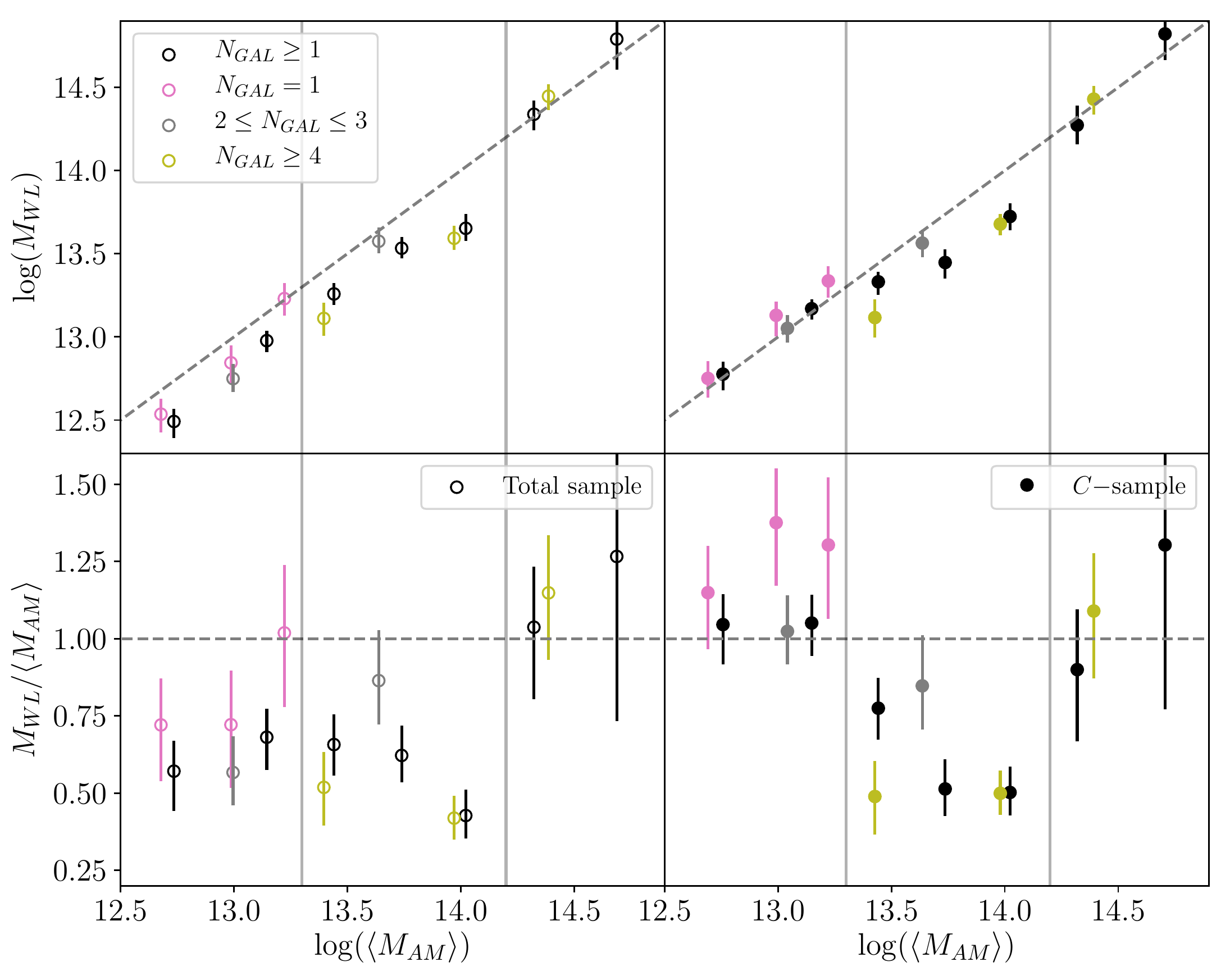}
    \caption{$M_\text{WL}$ lensing estimate (upper panels) and ratio between 
    both mass estimates (bottom panels)  vs. the average $M_\text{AM}$,
    for the different samples analysed detailed in Table \ref{table:Nresults}.
    In the left panel we show the results for the Total sample using open 
    circles and in the right panel for the $C-$sample with filled circles.
    Dashed grey line corresponds to the identity and vertical lines
    represent the limits for the low-, intermediate- and high-mass ranges. Masses are expressed in units
    of $\log{(h^{-1} M_\odot)}$}
    \label{fig:NMbins}
\end{figure*}

\begin{figure*}
    \centering
    \includegraphics[scale=0.55]{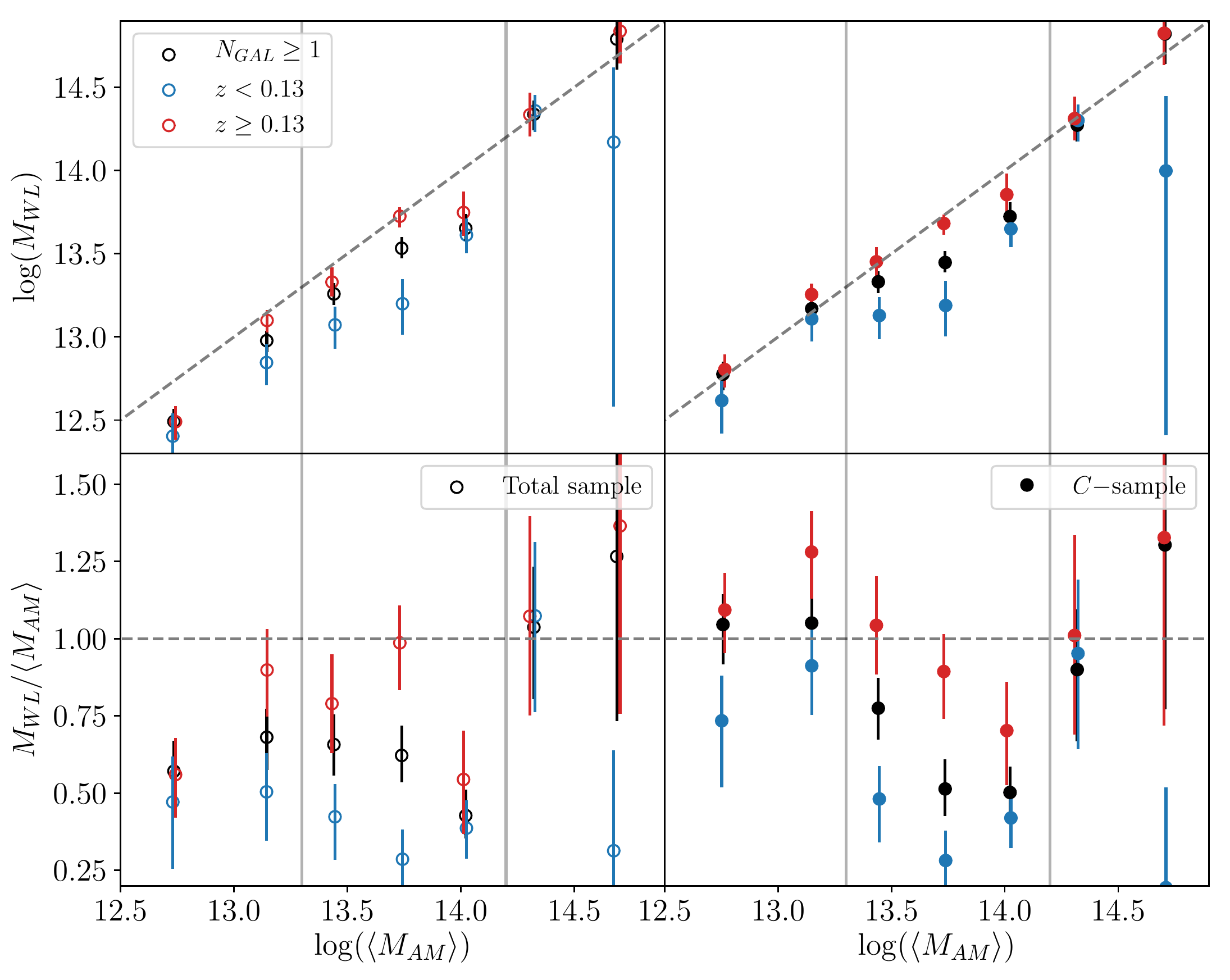}
    \caption{$M_\text{WL}$ lensing estimate (upper panels) and ratio between 
    both mass estimates (bottom panels)  vs. the average $M_\text{AM}$,
    for the subsamples selected according to the galaxy group redshift detailed in Table \ref{table:results}
    and the subsamples without richness restriction ($N_\text{GAL} \geq 1$) detailed in Table \ref{table:Nresults}.
    In the left panel we show the results for the Total sample using open 
    circles and in the right panel for the $C-$sample with filled circles.
    Dashed grey line corresponds to the identity  and vertical lines
    represent the limits for the low-, intermediate- and high-mass ranges. Masses are expressed in units
    of $\log{(h^{-1} M_\odot)}$}
    \label{fig:zMbins}
\end{figure*}

\begin{figure}
    \centering
    \includegraphics[scale=0.55]{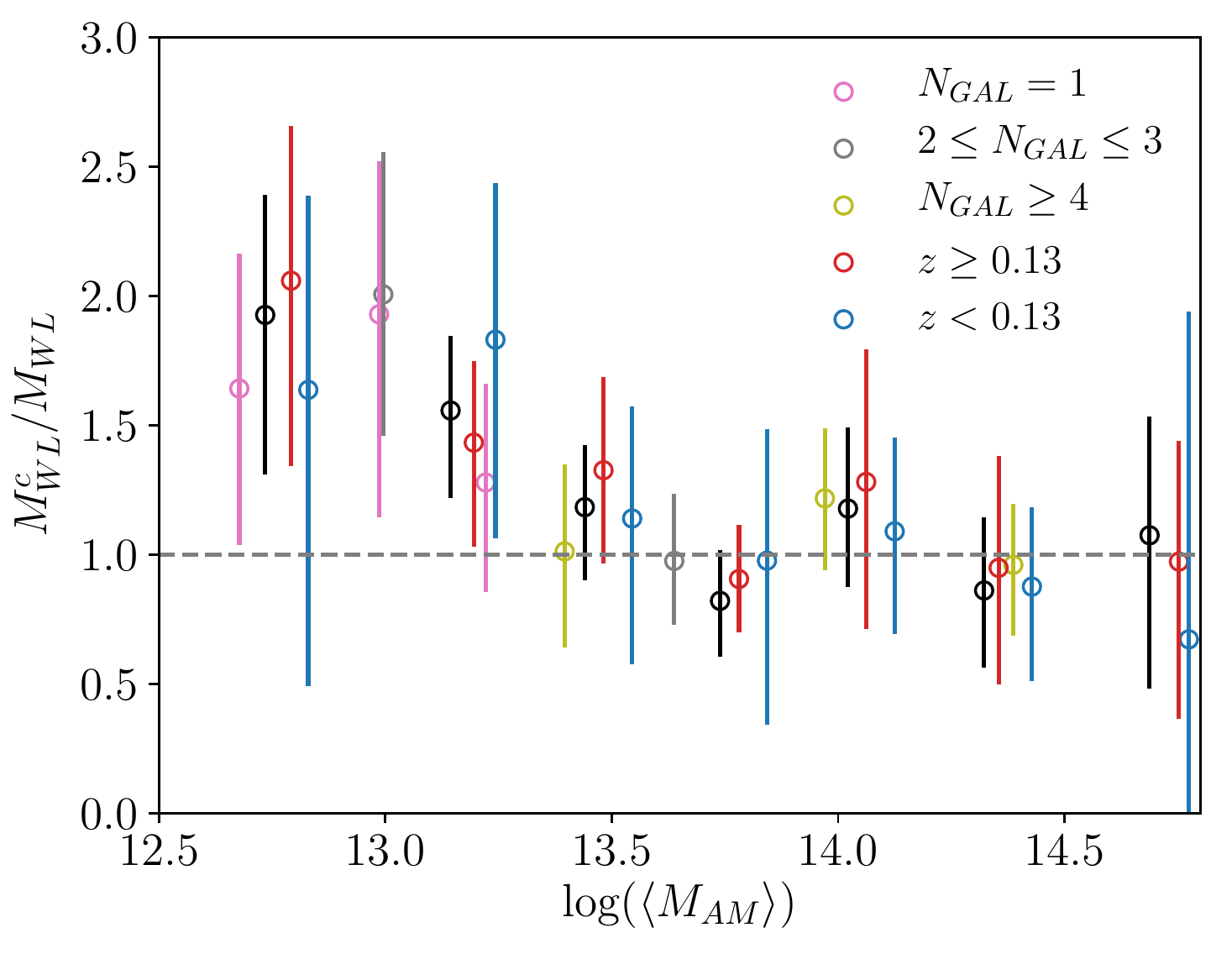}
    \caption{Mass ratio for the subsamples selected considering the
    whole sample of groups, $M_\text{WL}$, and the groups included in the 
    $C-$sample ($M^c_{WL}$), related to the average abundance matching 
    mass. Values for the $z \geq 0.13$ and $z < 0.13$ subsamples (red and blue dots) are
    shifted in the x-axis by $0.05$ and $0.10$ for a better visualisation.}
    \label{fig:Csample}
\end{figure}

In Figures \ref{fig:NMbins} and \ref{fig:zMbins} we show the relation between the average $M_\text{AM}$
and the lensing masses, $M_\text{WL}$, for the subsamples detailed in Tables \ref{table:Nresults}
and \ref{table:results}. In order to interpret the results we define three mass
regimes, the low-mass that includes groups with $\langle M_\text{AM} \rangle < 10^{13.3} h^{-1} M_\odot$,
the intermediate-mass that considers groups with $10^{13.3} h^{-1} M_\odot < \langle M_\text{AM} \rangle < 10^{14.2} h^{-1} M_\odot$
and the high-mass regime with groups that have $\langle M_\text{AM} \rangle > 10^{14.2} h^{-1} M_\odot$.

There is an excellent correlation between both mass estimates for all the 
subsamples considered. Nevertheless, for the subsamples selected in the whole redshift and richness range 
from the total sample of groups (empty black dots in both figures), lensing masses are systematically
underestimated for low- and intermediate-mass groups by a factor $\sim 0.6$. 
When considering the groups from the $C-$sample, we obtain a better agreement
between both mass estimates for the low-mass subsamples.
In Fig. \ref{fig:Csample} we compare lensing mass estimates
considering the total and $C-$sample. For the $C-$sample we expect higher lensing mass estimates since this selection proxy could improve the group selection, by considering systems with an elliptical brightest galaxy, since this morphological galaxy type is more related with denser environments. As it can be noticed, lensing estimates are
all in excellent agreement for all the groups with $\langle M_\text{AM} \rangle \geq 10^{13.3} h^{-1} M_\odot$.
On the other hand, for low-mass groups, lensing masses are about a factor two
higher when the BGM is an early-type galaxy. This is expected since the applied cut in the concentration index affects mainly
this mass range (see Fig.\ref{fig:hist}).

When considering a group richness binning to infer halo mass lensing estimates,
the subsamples that include groups with low-richness ($N_\text{GAL} \leq 3$) follow the 
trend found in the low-mass regime. Contrarily, at the intermediate-mass range, which
includes a richer range, $N_\text{GAL} \geq 4$, group masses show
lower lensing estimates compared to $\langle M_\text{AM} \rangle$. 
For this range of masses and richness we expect a higher contamination
by interlopers, since the purity of the group identification algorithm 
is lower at these ranges \citep[see Fig. 1 from][]{Rodriguez2020}.
This could bias the abundance matching masses to higher values since
a higher total luminosity is assigned. 

Finally, when the subsamples are selected according to the 
group redshift, we obtain systematically higher lensing masses
for the high redshift sample. Although both masses are 
in agreement within 1.5$\sigma$ for the low- and high-mass regimes, 
for intermediate-mass groups at higher redshift we obtain up to three times higher lensing masses
than for the groups located at lower redshift. These discrepancies
can be related with the observed differences when selecting the 
subsamples according to the richness, since intermediate-mass groups with $N_\text{GAL} \geq 4$
are mainly located at lower redshifts (see Fig.\ref{fig:hist}). 
In this mass range $75\%$ ($21\%$) 
of the groups located at $z < 0.13$ ($z \geq 0.13$) have $N_\text{GAL} \geq 4$.

\subsection{Relating lensing masses with the LOS velocity dispersion}
\label{subsec:zrelation}

\begin{figure}
    \centering
    \includegraphics[scale=0.55]{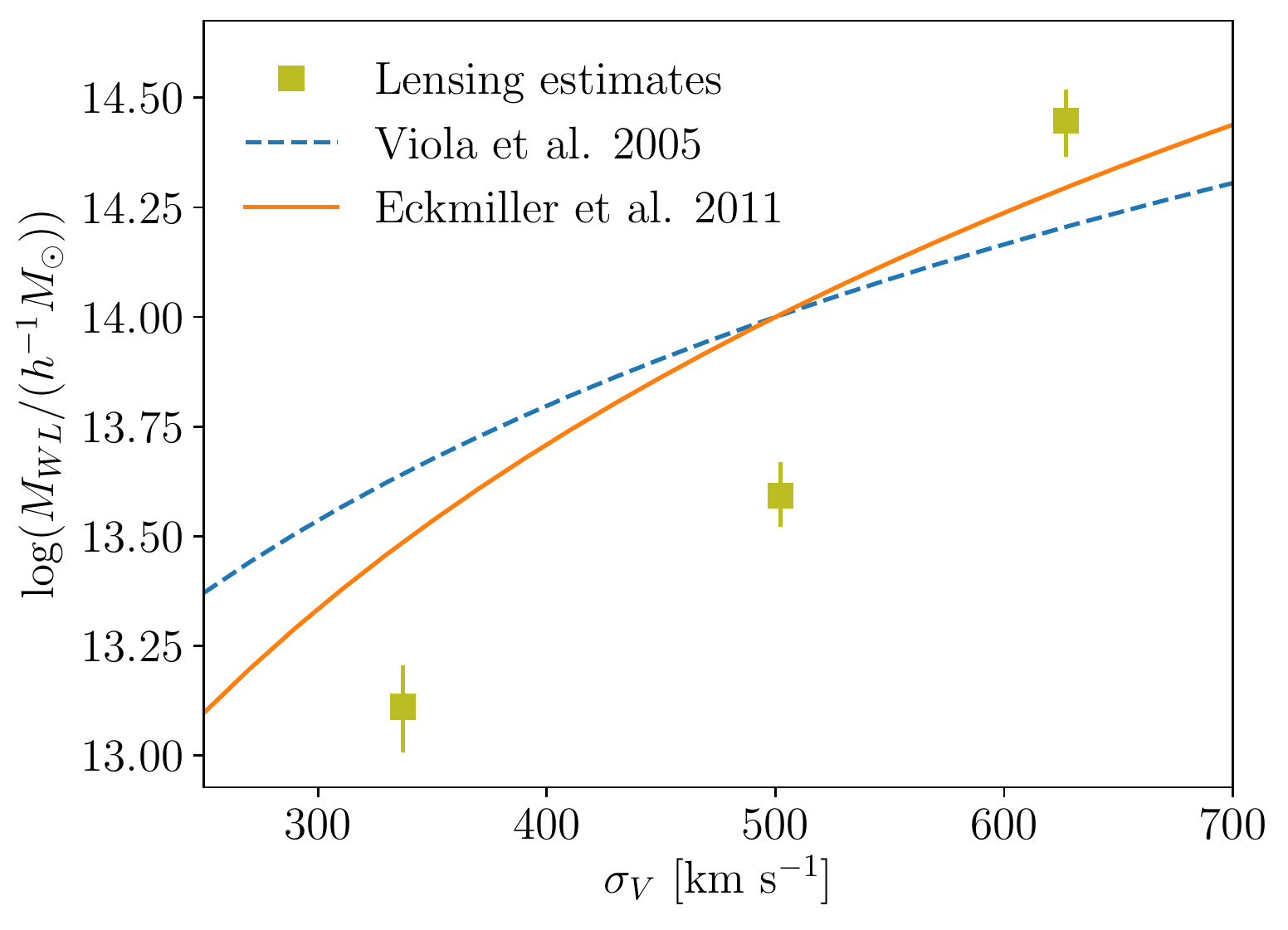}\\
    \caption{Comparison between the median LOS velocity dispersion, $\sigma_V$, 
    and derived lensing mass estimates, $M_\text{WL}$,  for groups
    with more than 4 members (yellow squares). Dashed blue line corresponds to the relation derived by \citet{Viola2015}
    between lensing mass estimates and the velocity dispersions for a sample of galaxy groups
    ($M_\text{WL}/(10^{14} h^{-1}M_\odot) = (\sigma_V/(500$ km\,s$^{-1}))^{1.89}$). The shaded blue
    region corresponds to the reported dispersion ($\sigma_{\log(M_\text{WL})} = 0.2$). In solid
    orange line we show the predicted relation according to numerical simulations \citep[$M_{200}/(10^{14} h^{-1}M_\odot) = (\sigma_V/(500 $ km\,s$^{-1}))^3$,][]{Eckmiller2011}}
    \label{fig:Mdyn}.
\end{figure}

Usually adopted mass estimates for low-richness galaxy systems are 
based on the dynamics of galaxies. These estimates are computed through  
spectroscopic redshifts and angular positions of galaxy members.
We have compared our derived lensing halo masses to
the median $\sigma_V$ provided in the 
catalogue for groups with $N_\text{GAL} \geq 4$.
(Fig. \ref{fig:Mdyn}). A good correlation is observed between these
parameters. According to numerical simulations,
a virial scaling relation of the form \citep[$M \propto \sigma^3$,][]{Evrard2008}
is expected between these parameters.
Nevertheless, a lower slope was found from previous weak lensing analysis
\citep[$M \propto \sigma^{\sim 2}$,][]{Han2015,Viola2015}. According to
\citet{Viola2015} the observed shallower mass-velocity relation
is mostly related to selection effects of the group sample. On the other hand,
our results (Fig. \ref{fig:Mdyn}) are more compatible with a steeper relation. 

It is important to highlight that $\sigma_V$ depend crucially on membership
assignment. The inclusion of interlopers might bias the velocity
dispersion to higher values. 
Moreover, virial mass estimates assume that the group/clusters are in dynamical
equilibrium. According to the results discussed in the previous
subsection, higher mass halos include a larger fraction of
miscentred groups. Thus, the highest mass bin sample may contain a larger fraction of merging 
systems. Another drawback about the dynamical estimates
relies on the simplicity of the model assumed to compute the masses, since the
relation between the projected velocity dispersion and the mass is held 
only up to the virial radius \citep{Alpaslan2012}. Nevertheless and in spite of these possible bias, the observed
good correlation between both parameters suggests that $\sigma_V$ provides
a suitable proxy for the mean halo mass.

\color{black}
\section{Summary and discussion}
\label{sec:summary}

In this work we have presented a weak lensing mass analysis of a galaxy group sample in the redshift range  $0.05 < z < 0.2$, identified using a combination of FOF and halo-based
methods. The group sample considered spans over a wide mass range, with $M_\text{WL}$ masses ranging from $10^{12.5}h^{-1}M_\odot$ to $10^{14.7}h^{-1}M_\odot$.
In order to explore the relation between the masses assigned according
to the abundance matching technique, $M_\text{AM}$, and the lensing estimates,
we split the total sample of groups in subsamples of
$\log(M_\text{AM})$ bins. We also considered different subsamples selected according to the group richness, $N_\text{GAL}$, and redshift.

For the lensing analysis we applied stacking techniques and combined four public lensing catalogues in order to improve the signal-to-noise ratio. Lensing profiles
were obtained by assuming that the brightest galaxy member
is located at the halo centre. We modelled the profiles 
by considering two free parameters, the fraction of well-centred groups, $p_\text{cc}$, and
the NFW mass, $M_\text{WL}$. 

According to a mock sample of groups identified in numerical simulations,
the fractions of well-centred groups derived from the lensing analysis
are in agreement with the ones expected due to uncertainties in the membership assignment.
Nevertheless, lensing estimates of the fraction of centred groups are mainly biased to lower values, specially for the higher mass subsamples. This result may be due to the inclusion of a larger
fraction of merging systems in these subsamples.  

Lensing masses obtained for the different subsamples of groups considered correlate well with the average abundance matching estimates. This result provides observational evidence of a tight correlation between the halo mass and the characteristic group luminosity. Therefore, it supports the use of the group luminosity as a mass proxy, specially for low massive systems.
However, masses based on the group luminosity tend to predict higher values than the determined by the weak-lensing analysis,
for low- and intermediate-mass groups ($\langle M_\text{AM} \rangle < 10^{14.2} h^{-1} M_\odot$).
When considering only the groups with an early-type central galaxy selected according
to the concentration index, the agreement between lensing masses and $M_\text{AM}$
improves significantly in the low-mass regime ($\langle M_\text{AM} \rangle < 10^{13.3} h^{-1} M_\odot$).
This behaviour is observed for all the subsamples included in this mass range, regardless 
of the richness and redshift group. Since it is expected that early-type galaxies 
are associated with denser environments \citep{Dressler1980,Postman1984,Balogh2004,Kauffmann2004},
considering this proxy for the group selection could improve the identification. 
Also, a higher dispersion of the luminosity-halo mass relation is expected for this mass range \citep{Yang2005}. It is important to take into account that the abundance matching relies on an 
oversimplified one-to-one relation between the characteristic luminosity of each group
and the halo mass, which neglects the effects of possible biases introduced by other properties
such as the morphology or colours of the member galaxies. 

For intermediate-mass groups ($10^{13.3} h^{-1} M_\odot < \langle M_\text{AM} \rangle < 10^{14.2} h^{-1} M_\odot$) lensing masses are systematically biased to lower values
for all the subsamples considered. For this mass range, we also find masses to be biased towards lower values for groups at lower redshifts ($z < 0.13$). When splitting the subsamples according to 
group richness, the bias in this mass range prevails only for the groups with more than
three members. Since at this richness and mass range we expect higher uncertainties in membership 
assignment, it can be argued that a significant inclusion of
interlopers are affecting the characteristic luminosity assigned. This could
also explain the observed bias in the low-redshift subsamples since
at this mass and redshift range, $75\%$ of the groups have $N_\text{GAL} \geq 4$. 

Finally, for the 
high-mass groups ($\langle M_\text{AM} \rangle > 10^{14.2} h^{-1} M_\odot$), we obtain
a good agreement between mass estimates for all the considered subsamples. 
This is in agreement with a more constrained relation between the group luminosity 
and the halo mass for the systems with higher masses \citep{Kang2005}, favouring the
one-to-one relation in which the abundance matching mass is based. 

In addition with a possible bias introduced by interlopers in the galaxy group identification, the observed discrepancies between predicted masses based on the group luminosity and the derived according to the lensing study, can be also related with the intrinsic scatter between the luminosity and the halo mass. A deep inspection between possible bias introduced in the mass assignment according to the group characteristic luminosity using hydrodynamic simulations can help to asses the observed differences. On the other hand, the analysis of a larger group sample which will increase the lensing signal, thus allowing an improvement in the modelling of the profiles, can also provide better constrained lensing masses to discard possible bias introduced in the study.

We have also compared our lensing masses to the median LOS
velocity dispersion of the subsamples of groups with more than four members.
As for the abundance matching mass comparison, lensing masses
for groups in the intermediate-mass range are biased to lower values, 
compared with the median velocity dispersion predicted by simulations. 
Once again, the inclusion of interlopers might be biasing the observed LOS 
velocity dispersion, $\sigma_V$, to higher values. We highlight
that the derived good correlation between both parameters indicates that $\sigma_V$
also provides a good proxy for the halo masses, but its limited to systems with more than four members.

The results derived by the analyses presented in this work,
can serve as important tests for the mass-proxy estimates
in a wide mass range of galaxy systems. A well calibrated
mass-proxy that can constrain the mean halo masses is important
in order to better characterise galaxy systems and to 
use them as cosmological probes. Although there is still a long
way ahead in order to quantify the possible
biases introduced, this work supports the use of abundance matching
techniques for mass estimates of diverse samples of galaxy systems.

\section*{Acknowledgements}

We are highly thankful to the anonymous referee for their useful comments, which helped to improve this paper. This work was partially supported by the Consejo Nacional
de  Investigaciones  Cient\'ificas  y  T\'ecnicas  (CONICET, Argentina), the Secretar\'ia de Ciencia y Tecnolog\'ia de la Universidad Nacional de C\'ordoba (SeCyT-UNC, Argentina), the Brazilian Council for Scientific and Technological Development (CNPq) and the Rio de Janeiro Research Foundation (FAPERJ). 
We acknowledge the PCI BEV fellowship program from
MCTI and CBPF.
MM acknowledges FAPERJ and CNPq for financial support. FORA BOZO.
This  work  is  based  on  observations  obtained  with
MegaPrime/MegaCam,  a  joint  project  of  CFHT  and
CEA/DAPNIA, at the Canada--France--Hawaii Telescope
(CFHT), which is operated by the National Research
Council (NRC) of Canada, the Institut National des Sciences de l'Univers of the Centre National de la Recherche Scientifique (CNRS) of France, and the University of
Hawaii. The Brazilian partnership on CFHT is managed
by the Laborat\'orio Nacional de Astrof \'isica (LNA). We
thank the support of the Laborat\'orio Interinstitucional de
e-Astronomia (LIneA). We thank the CFHTLenS team
for their pipeline development and verification upon which
much of the CS82 survey pipeline was built.\\
This research used the facilities of the
Canadian Astronomy Data Centre operated by the National Research
Council of Canada with the support of the Canadian Space Agency.
RCSLenS data processing was made possible thanks to significant
computing support from the NSERC Research Tools and Instruments grant
program.
Based on data products from observations made with ESO Telescopes at the La Silla Paranal Observatory under programme IDs 177.A-3016, 177.A-3017 and 177.A-3018. 

\section*{Data Availability}

The datasets were derived from sources in the public domain: CFHTLenS (http://www.cadc-ccda.hia-iha.nrc-cnrc.gc.ca/en/community/CFHTLens), RCSLenS (https://www.cadc-ccda.hia-iha.nrc-cnrc.gc.ca/en/community/rcslens), KiDS-450 (http://kids.strw.leidenuniv.nl/cosmicshear2018.php),
redMaPPer (http://risa.stanford.edu/redmapper/). CS82 can be accessed on request by emailing
to martin\@cbpf.br. The data derived in this article are available on request to the corresponding author.



\bibliographystyle{mnras}
\bibliography{references} 


\appendix

\section{Mass estimates derived for the individual lensing surveys}
\label{app:test}
\begin{figure}
    \centering
    \includegraphics[scale=0.6]{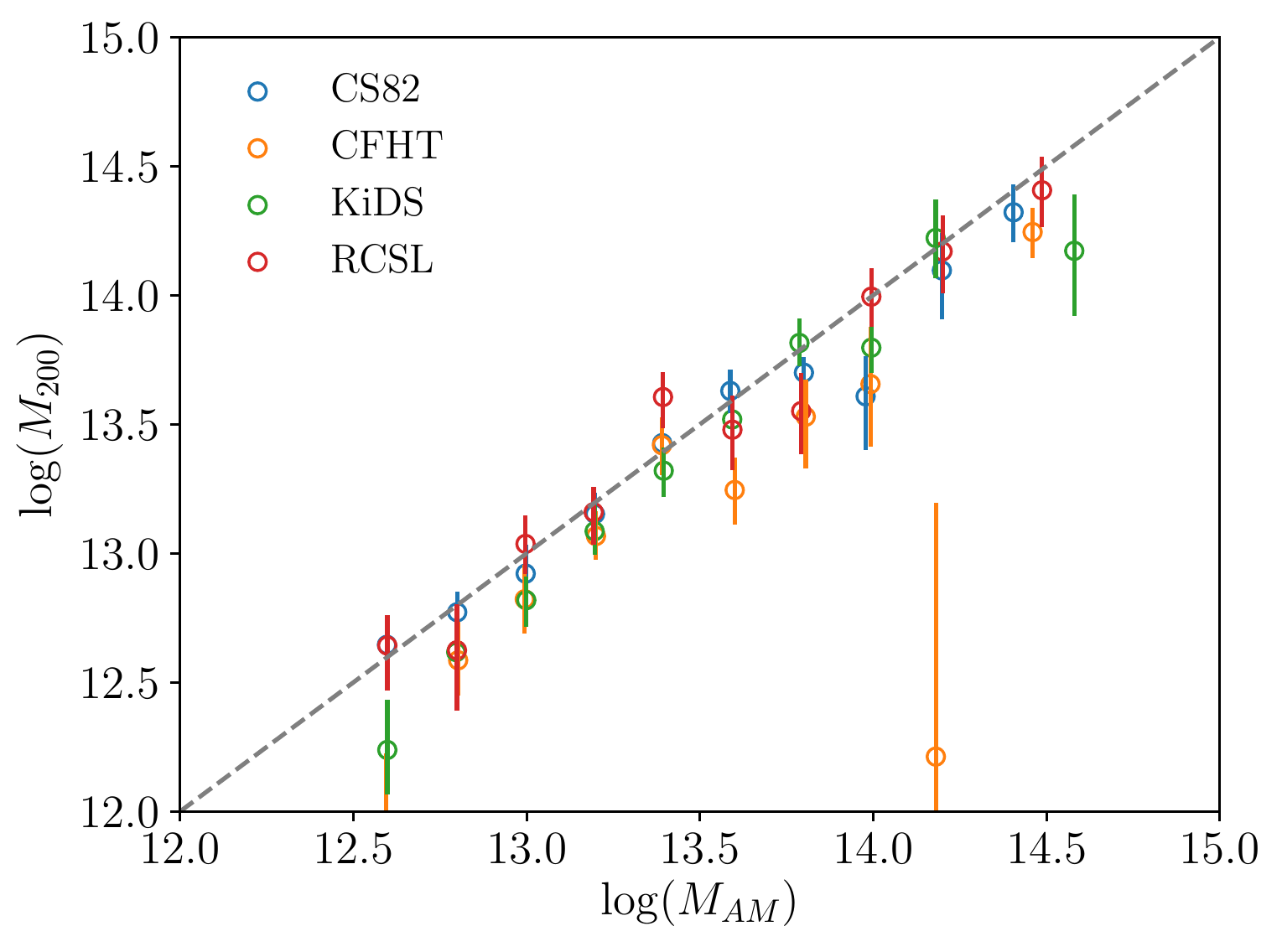}
    \caption{Fitted lensing masses derived using the individual shear data sets  vs. the average $M_\text{AM}$.}
    \label{fig:test}
\end{figure}

In order to test the combination of the shear catalogues used for the analysis and presented in Sec. \ref{sec:data}, we derive the $M_\text{WL}$ by fitting the profiles computed using the individual lensing catalogs combined in this work (CFHTLenS, CS82, RCSLens,KiDS-450). Profiles were obtained by selecting the groups from the total group sample ($N_\text{GAL} \geq 1$) according to the $\log M_\text{AM}$ bins specified in Table \ref{table:Nresults}. In Table \ref{tab:test} we show the number of groups considered for the stacking in each bin and lensing survey. In Fig. \ref{fig:test} we plot the relation between the mean $M_\text{AM}$ and lensing estimates using the individual surveys. According to this comparison, not significant biases are observed in the computed masses. 

\begin{table}
    \centering
    \begin{tabular}{c c c c c}
    \hline
    \hline
    $\log{M_\text{AM}}$ & CFHT  & CS82   &  KiDS-450 & RCSLens \\
    \hline
$[12.5,13.0)$ & 1405 &  2848 &  5412   &  4481  \\
$[13.0,13.3)$ &  322 &  729  &  1295   &  1132  \\
$[13.3,13.6)$ &  148 &  337  &   581   &   490  \\
$[13.6,13.9)$ &   64 &  132  &   244   &   200  \\
$[13.9,14.2)$ &   24 &   50  &   105   &    85  \\
$[14.2,14.5)$ &    7 &   21  &    20   &    33  \\
$[14.5,15.0)$ &    2 &    0  &     4   &     3  \\
\end{tabular}
    \caption{Columns: (1) $\log{M_\text{AM}}$ bins; (2), (3), (4) and (5) Number of groups in each bin considered for the stacking using the correspondent lensing catalog}
    \label{tab:test}
\end{table}

\section{2D probability distributions}
\label{app:corner}

\begin{figure*}
    \centering
    \includegraphics[scale=0.6]{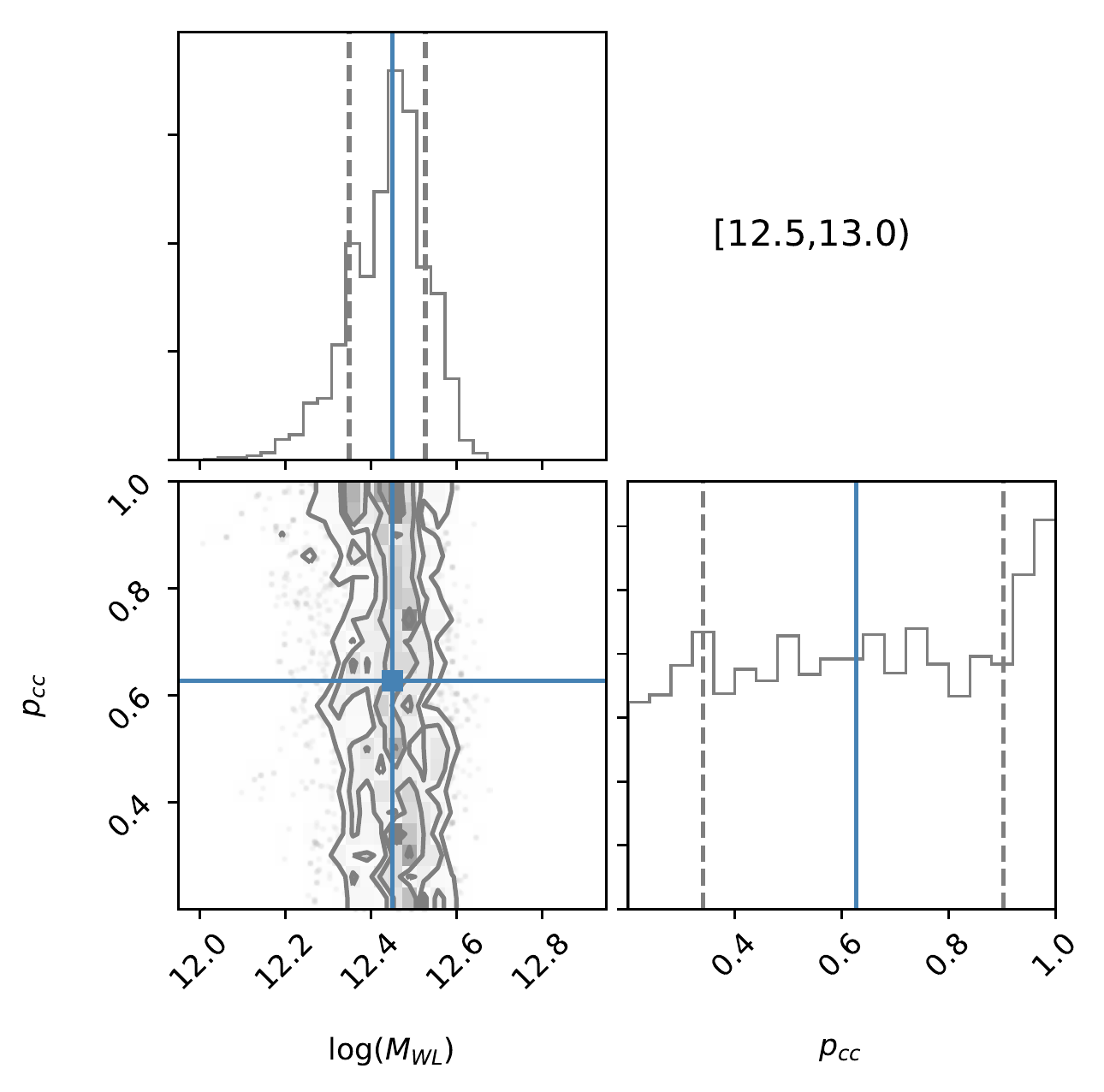}
    \includegraphics[scale=0.6]{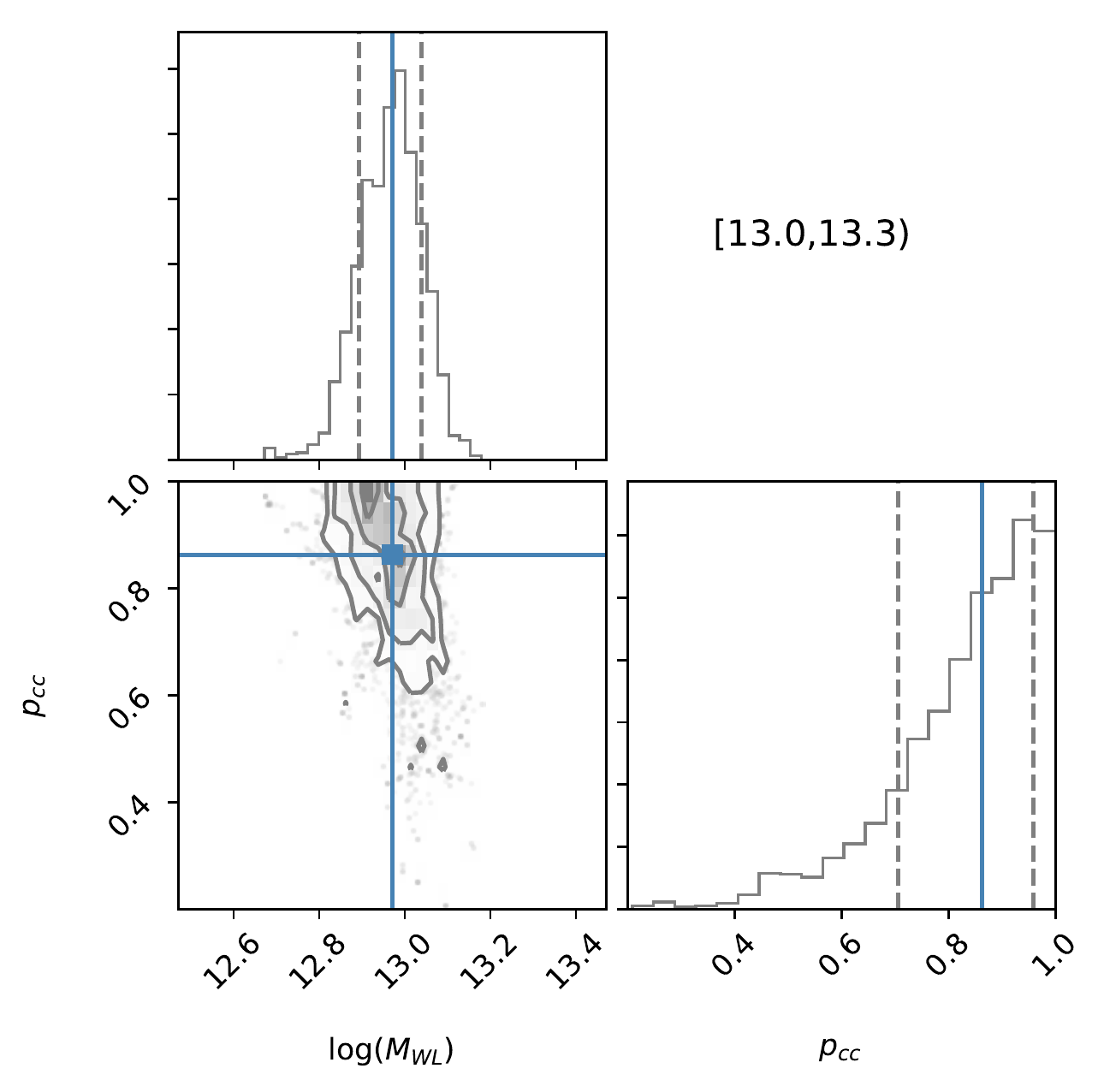}
    \includegraphics[scale=0.6]{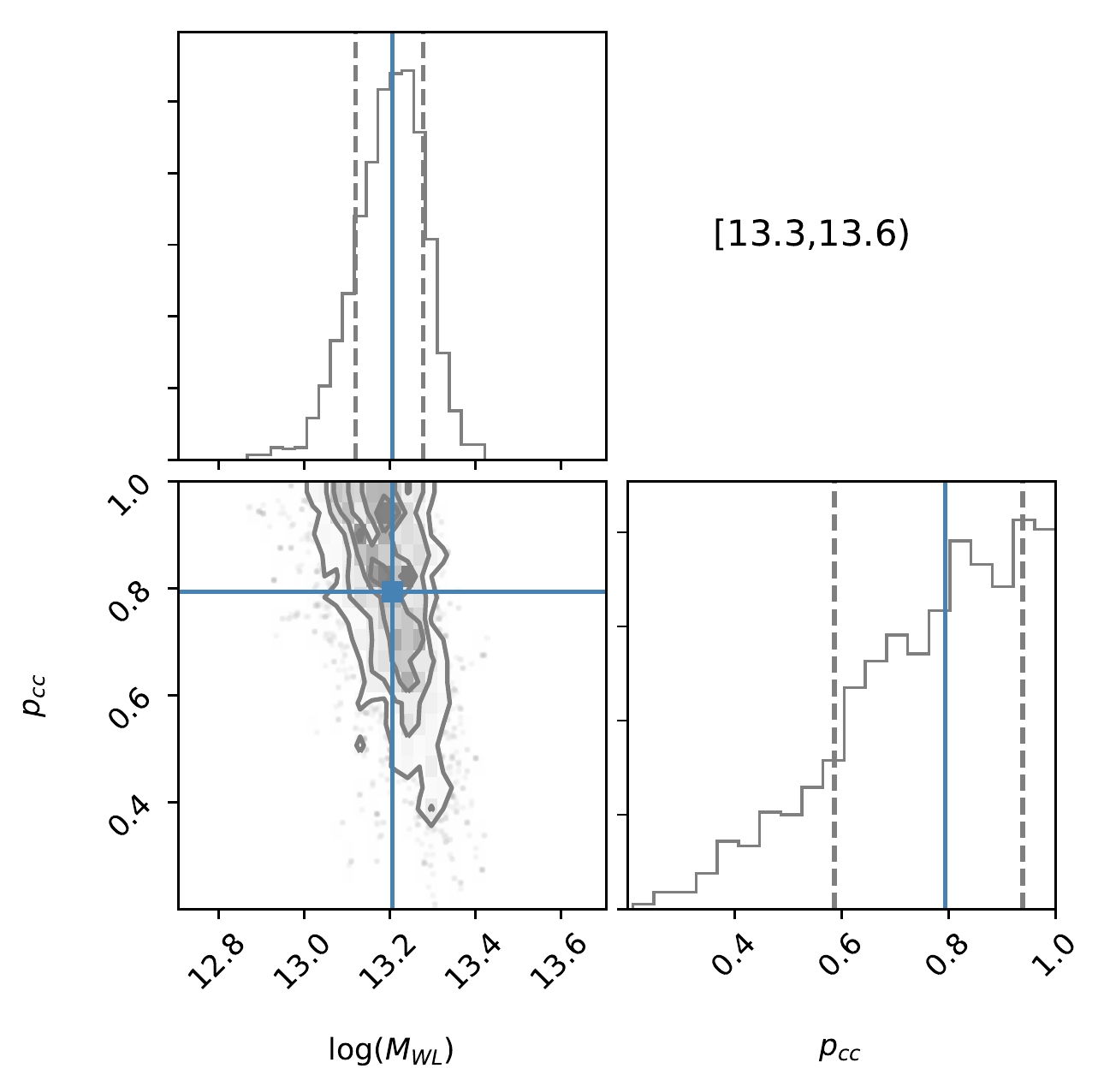}
    \includegraphics[scale=0.6]{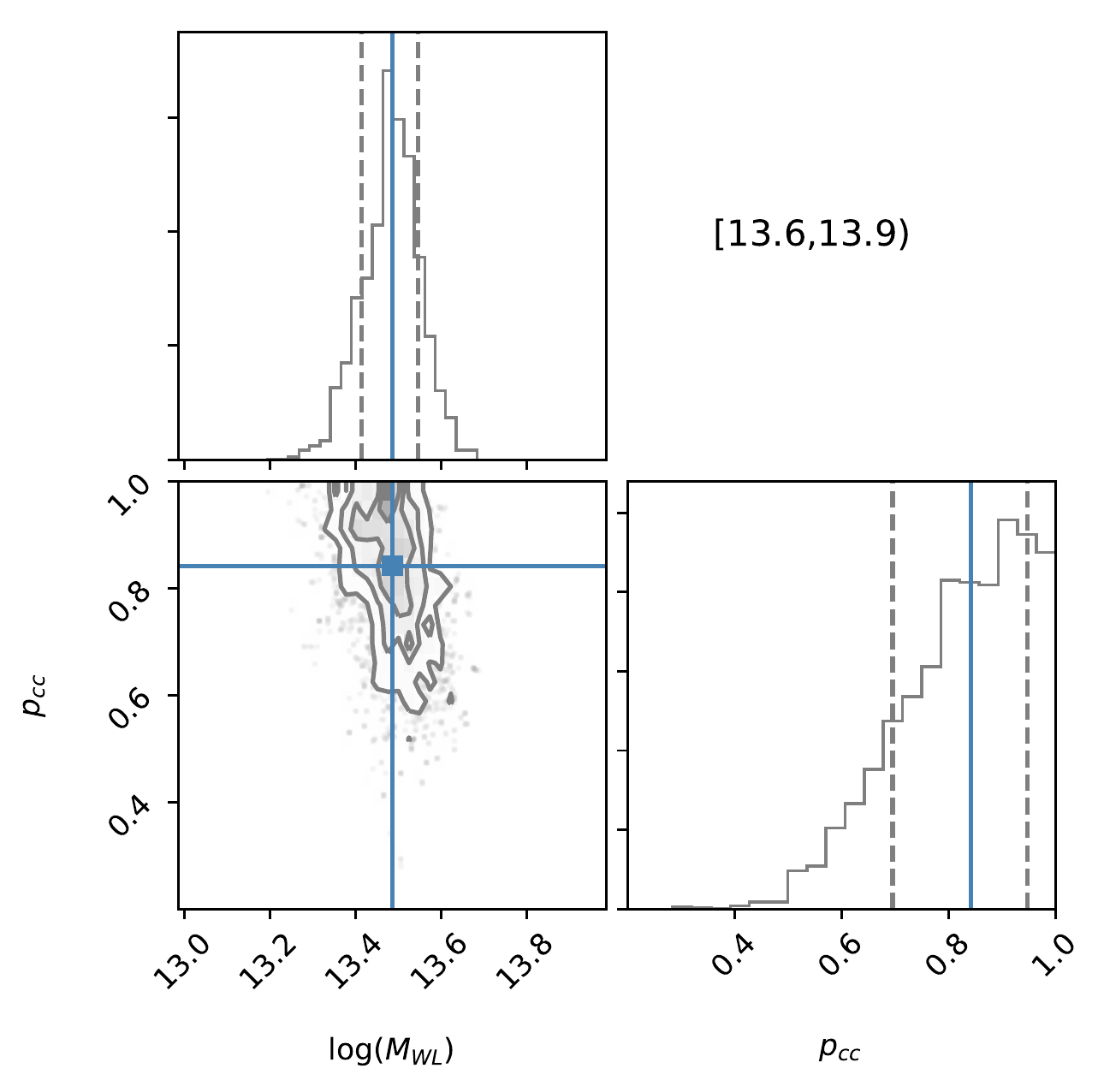}
    \caption{One and two dimensional projections of the posterior probability distributions of the fitted parameters, $\log(M_\text{WL})$ and $p_\text{cc}$, for the first four bins described in Table \ref{table:Nresults}. Solid line represents the adopted median value while dashed lines correspond to the 16-th and 84-th percentiles.}
    \label{fig:corner1}
\end{figure*}

\begin{figure*}
    \centering
    \includegraphics[scale=0.6]{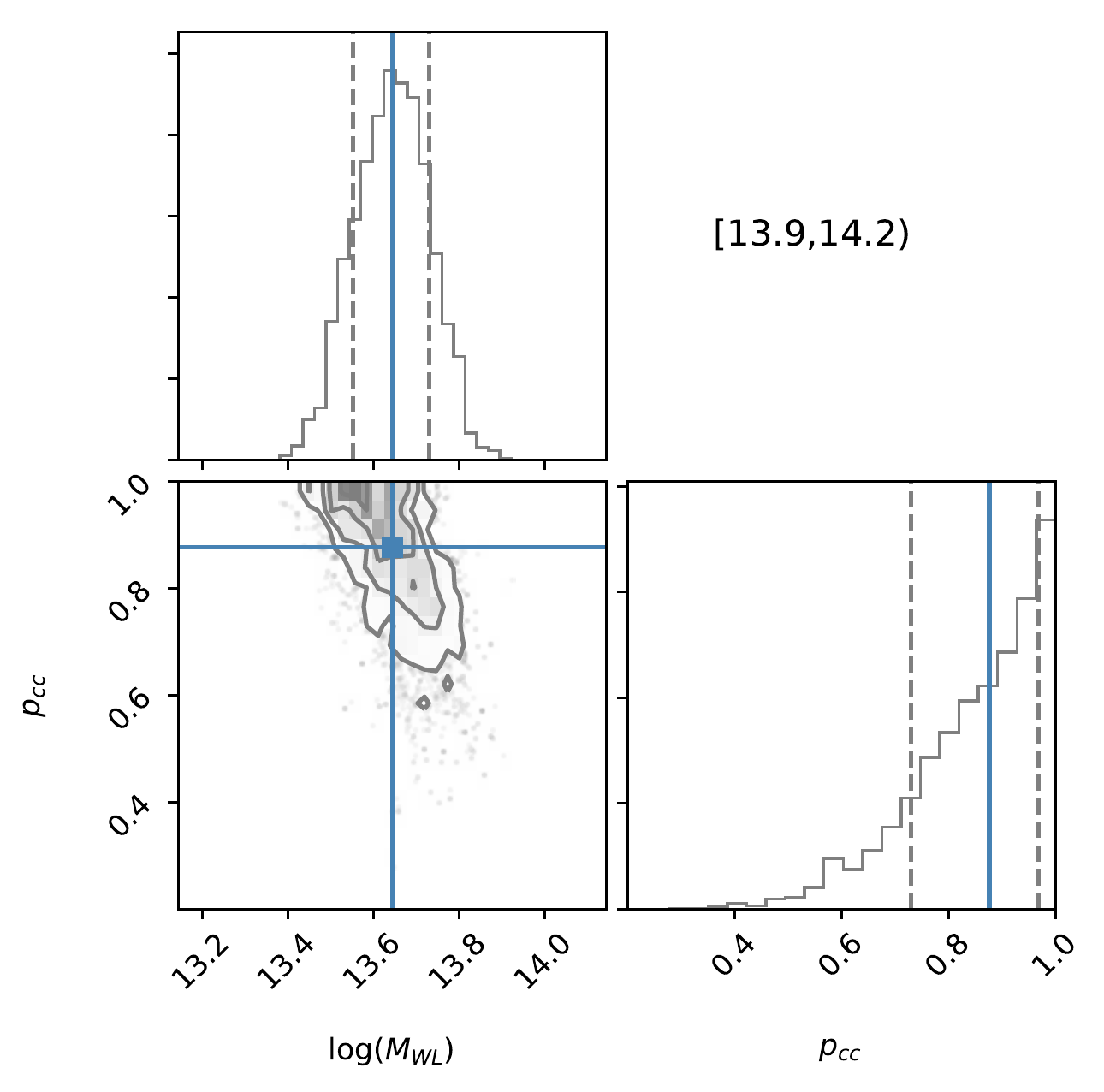}
    \includegraphics[scale=0.6]{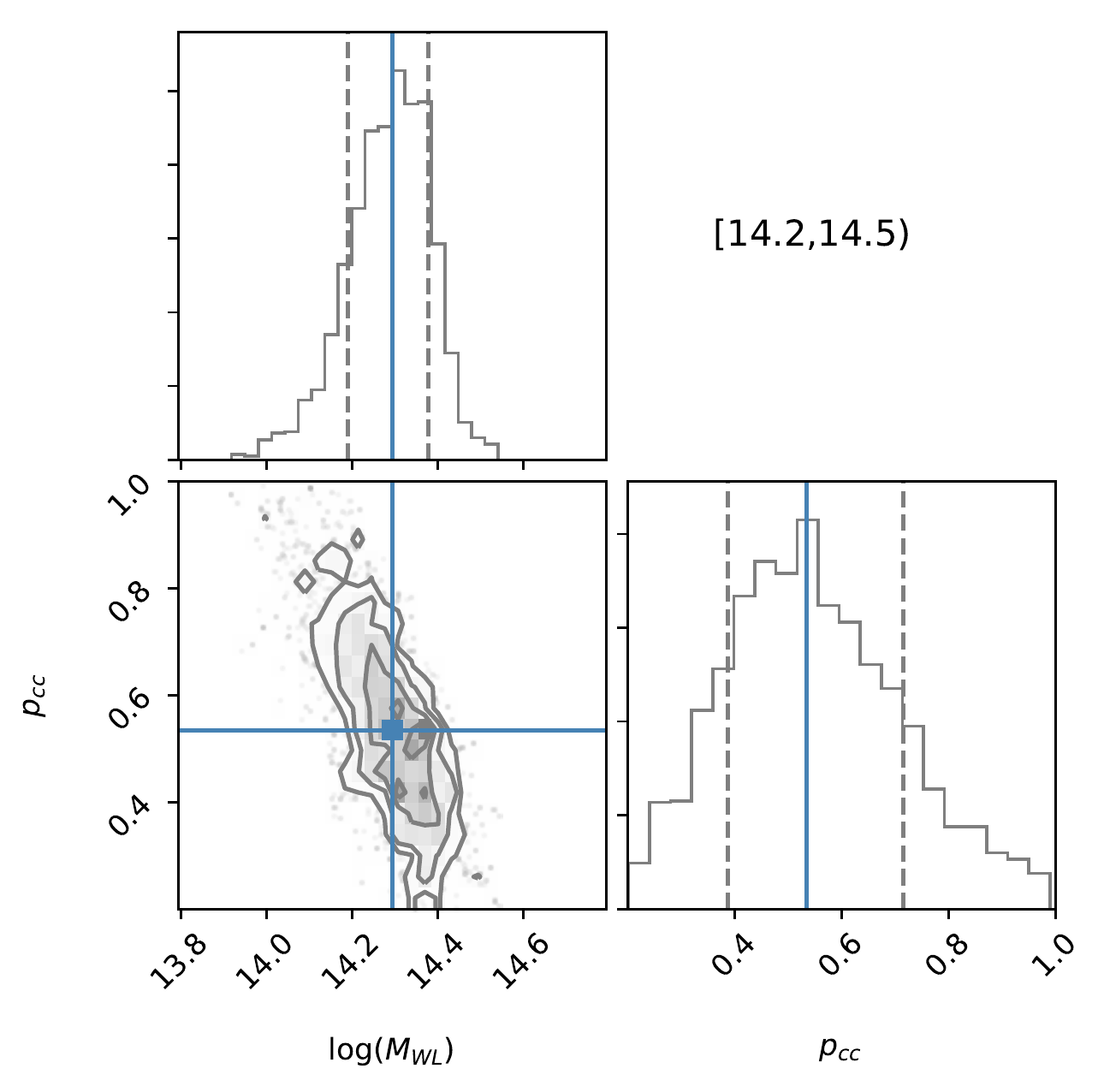}
    \includegraphics[scale=0.6]{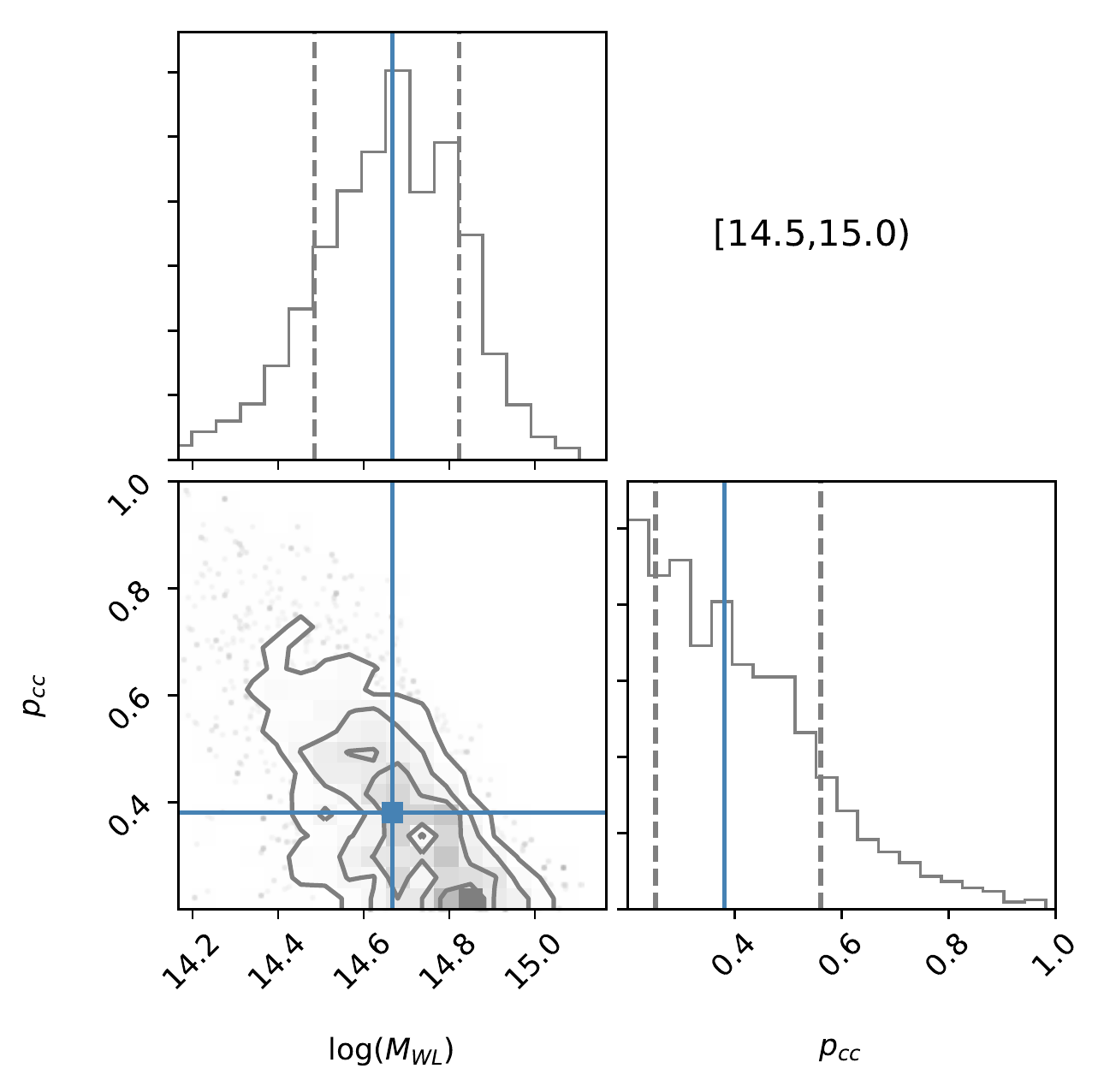}
\caption{One and two dimensional projections of the posterior probability distributions of the fitted parameters, $\log(M_\text{WL})$ and $p_\text{cc}$, for the latest three bins described in Table \ref{table:Nresults}. Solid line represents the adopted median value while dashed lines correspond to the 16-th and 84-th percentiles.}
    \label{fig:corner2}
\end{figure*}

We show in Fig. \ref{fig:corner1} and \ref{fig:corner2} the posterior distribution of the fitted parameters $\log(M_\text{WL})$ and $p_\text{cc}$ for the total sample binned according the described bins in Table \ref{table:Nresults}, with no restriction in richness ($N_\text{GAL} \geq 1$).

\section{Luminosity distributions}
\label{app:lum}

\begin{figure*}
    \centering
    \includegraphics[scale=0.6]{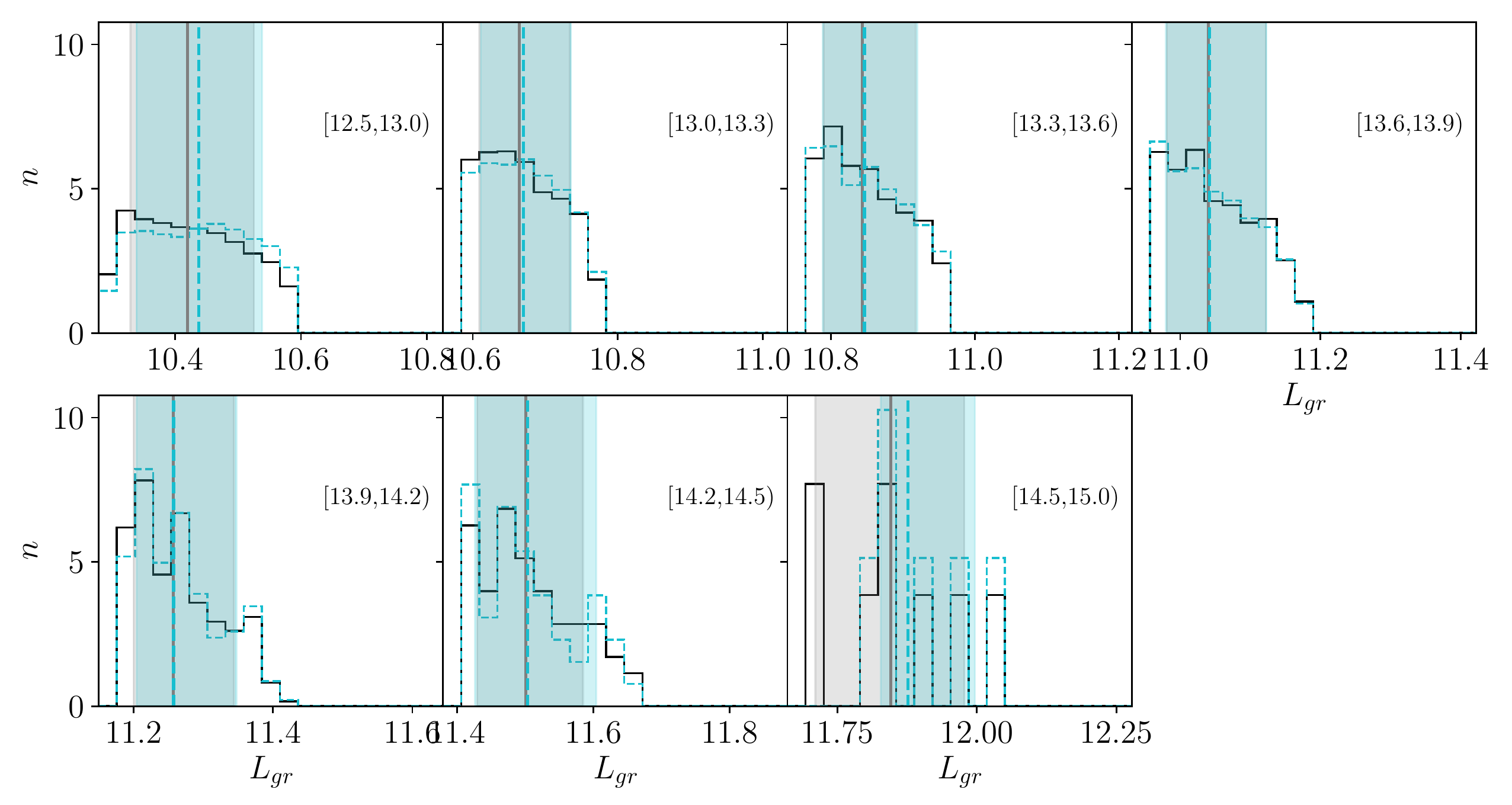}
    \caption{Normalised characteristic luminosity group distributions, $L_\text{gr}$, for the total (black solid line) and $C-$samples (cyan dashed line) selected in the $\log M_\text{AM}$ ranges indicated in each panel. Vertical lines correspond to the median values and the shadow region enclose 15- and 85-th percentiles. These samples correspond to the first 7 rows described in Table \ref{table:Nlum}, in the whole richness and redshift range ($N_\text{GAL} \geq 1$ and $0.05 \leq z < 0.2$). }
    \label{fig:distlum}
\end{figure*}

\begin{figure*}
    \centering
    \includegraphics[scale=0.6]{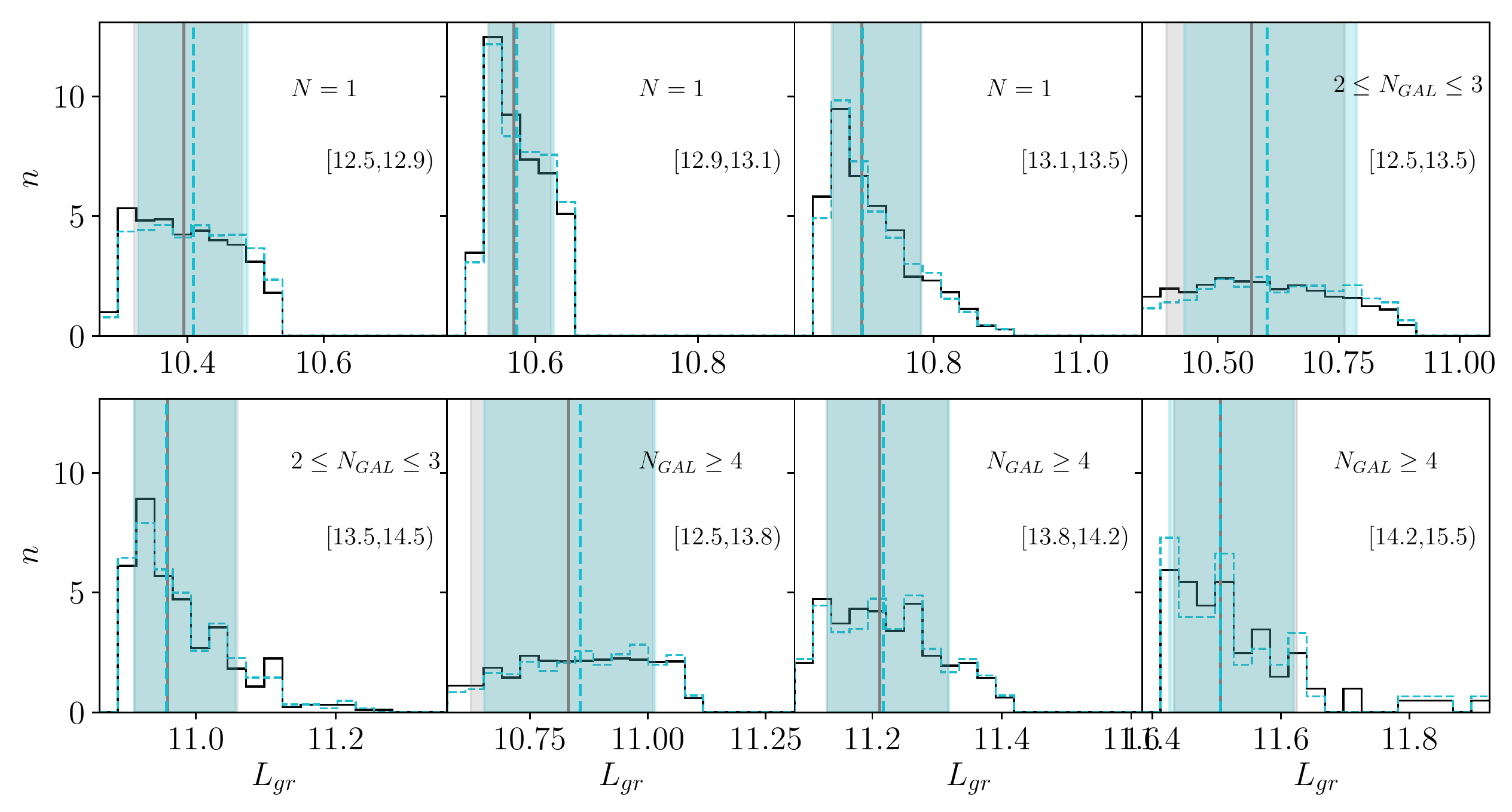}
    \caption{Normalised characteristic luminosity group distributions, $L_\text{gr}$, for the total (black solid line) and $C-$samples (cyan dashed line) selected in the $\log M_\text{AM}$ ranges indicated in each panel. Vertical lines correspond to the median values and the shadow region enclose 15- and 85-th percentiles. These samples correspond to the last 8 rows described in Table \ref{table:Nlum}, in the whole richness and redshift range ($N_\text{GAL} \geq 1$ and $0.05 \leq z < 0.2$) and in the richness range indicated in each panel. }
    \label{fig:distlumN}
\end{figure*}

\begin{figure*}
    \centering
    \includegraphics[scale=0.6]{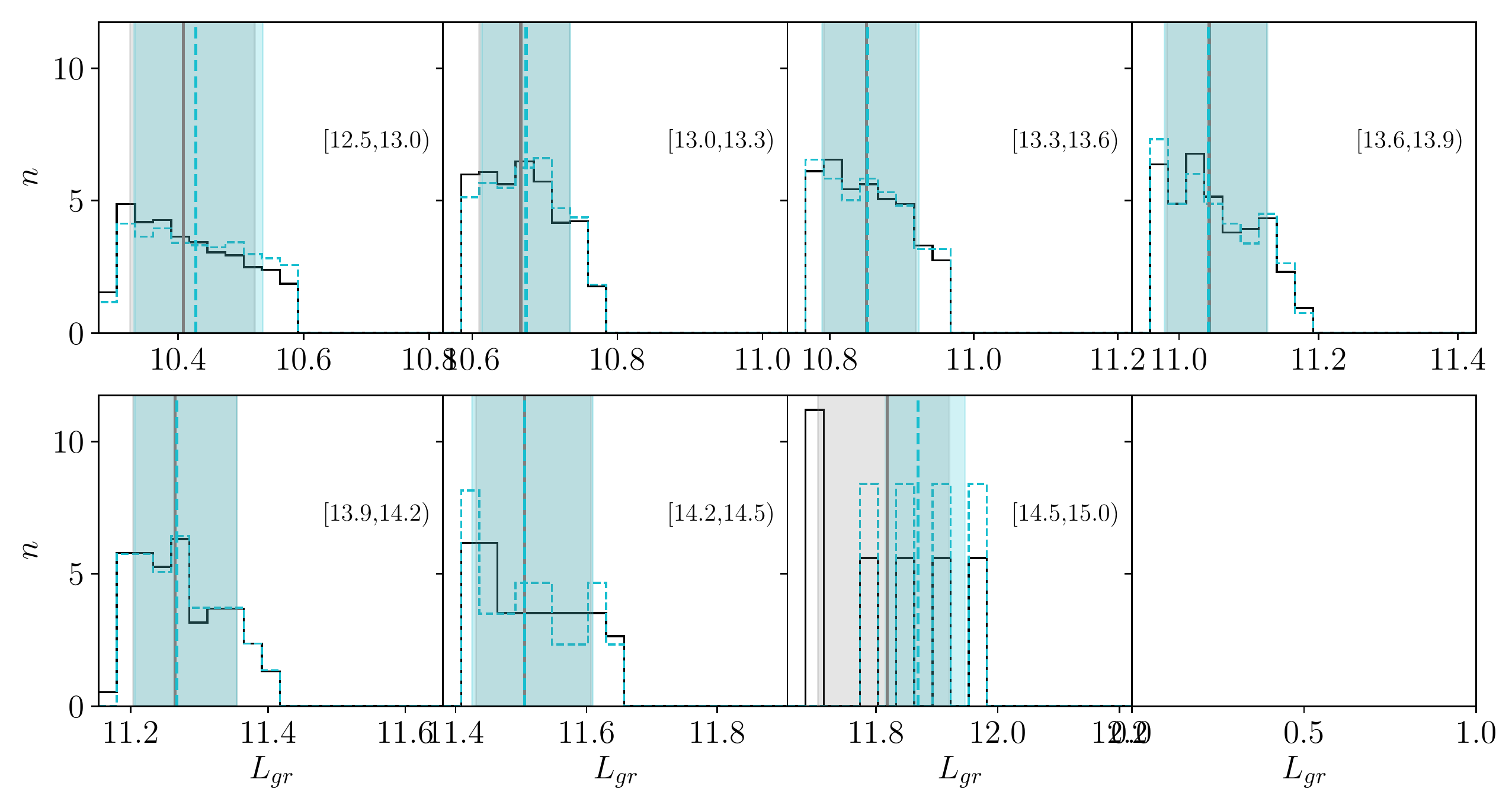}
    \caption{Normalised characteristic luminosity group distributions, $L_\text{gr}$, for the total (black solid line) and $C-$samples (cyan dashed line) selected in the $\log M_\text{AM}$ ranges indicated in each panel. Vertical lines correspond to the median values and the shadow region enclose 15- and 85-th percentiles. These samples correspond to the first 7 rows described in Table \ref{table:zlum}, in the whole richness range and with $0.05 \leq z < 0.13$. }
    \label{fig:distlumzL}
\end{figure*}

\begin{figure*}
    \centering
    \includegraphics[scale=0.6]{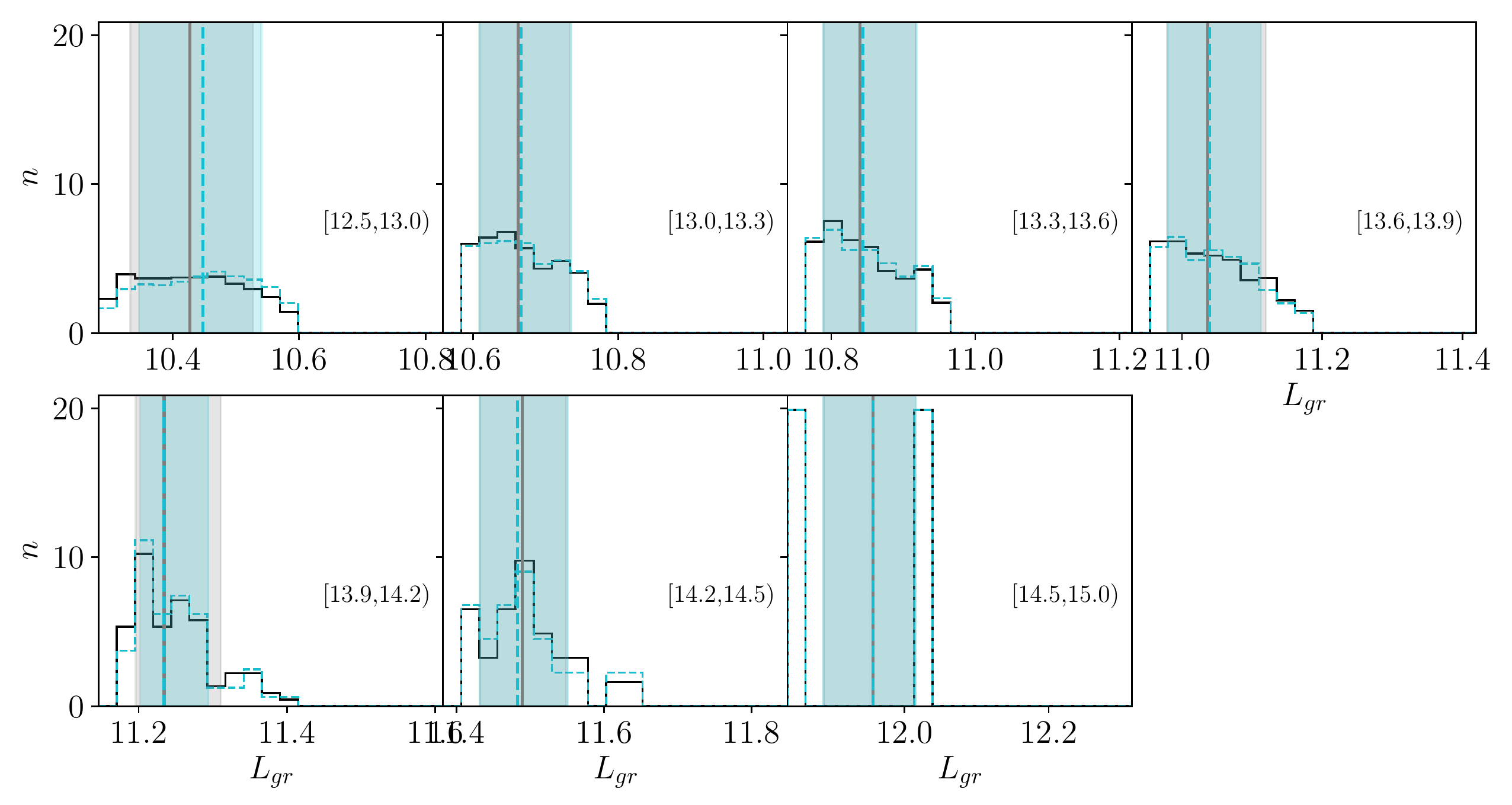}
    \caption{Normalised characteristic luminosity group distributions, $L_\text{gr}$, for the total (black solid line) and $C-$samples (cyan dashed line) selected in the $\log M_\text{AM}$ ranges indicated in each panel. Vertical lines correspond to the median values and the shadow region enclose 15- and 85-th percentiles. These samples correspond to the latest 7 rows described in Table \ref{table:zlum}, in the whole richness range and with $0.13 \leq z < 0.2$. }
    \label{fig:distlumzH}
\end{figure*}

In this appendix we detail the characteristic luminosity distributions, $L_\text{gr}$, for each subsample considered for the stacking in the lensing analysis. We show the distributions in the Figures \ref{fig:distlum}, \ref{fig:distlumN}, \ref{fig:distlumzL} and \ref{fig:distlumzH} for the total sample and $C-$sample selected according to the assigned abundance matching mass, richness and redshift. In Tables \ref{table:Nlum} and \ref{table:zlum} we give the median values and 16- and 84-th percentiles of each selected subsample of groups.

\begin{table*}
\begingroup
\setlength{\tabcolsep}{10pt} 
\renewcommand{\arraystretch}{1.5} 
\centering
\caption{Fitted parameters for the analysed galaxy groups in the whole redshift range ($0.05 \leq z < 0.2$).}
\begin{tabular}{c c | c c c | c c c}
\hline
\hline
\rule{0pt}{1.05em}%
Richness  & $\log{M_\text{AM}}$ & \multicolumn{3}{c|}{Total sample} & \multicolumn{3}{c}{$C-$sample} \\
selection       &                                   & $L_{50}$ & $L_{15}$ & $L_{85}$ & $L_{50}$ & $L_{15}$ & $L_{85}$  \\
\hline
\rule{0pt}{1.05em}   
$N_\text{GAL} \geq 1$ & $[12.5,13.0)$ & 10.42 & 10.33 & 10.53 & 10.44 & 10.34 & 10.54  \\ 
                      & $[13.0,13.3)$ & 10.66 & 10.61 & 10.73 & 10.67 & 10.61 & 10.74  \\ 
                      & $[13.3,13.6)$ & 10.84 & 10.79 & 10.92 & 10.85 & 10.79 & 10.92  \\
                      & $[13.6,13.9)$ & 11.04 & 10.98 & 11.12 & 11.04 & 10.98 & 11.12  \\
                      & $[13.9,14.2)$ & 11.26 & 11.20 & 11.34 & 11.26 & 11.20 & 11.35  \\
                      & $[14.2,14.5)$ & 11.50 & 11.43 & 11.58 & 11.50 & 11.43 & 11.60  \\
                      & $[14.5,15.0)$ & 11.85 & 11.71 & 11.98 & 11.88 & 11.83 & 12.00  \\
\hline
  $N_\text{GAL} =1$   & $[12.5,12.9)$ & 10.40 & 10.32 & 10.48 & 10.41 & 10.33 & 10.49  \\ 
                      & $[12.9,13.1)$ & 10.57 & 10.54 & 10.62 & 10.58 & 10.54 & 10.62  \\ 
                      & $[13.1,13.5)$ & 10.70 & 10.66 & 10.78 & 10.70 & 10.66 & 10.78  \\
\hline                           
$2 \leq N_\text{GAL} \leq 3$  & $[12.5,13.5)$ & 10.57 & 10.39 & 10.76 & 10.60 & 10.43 & 10.79  \\
                              & $[13.5,14.5)$ & 10.96 & 10.91 & 11.06 & 10.96 & 10.91 & 11.06  \\
\hline                           
  $N_\text{GAL} \geq 4$      & $[12.5,13.8)$ & 10.83 & 10.62 & 11.01 & 10.86 & 10.65 & 11.02  \\
                             & $[13.8,14.2)$ & 11.21 & 11.13 & 11.32 & 11.22 & 11.13 & 11.32  \\
                             & $[14.2,15.5)$ & 11.51 & 11.43 & 11.62 & 11.51 & 11.43 & 11.62  \\

\hline         
\end{tabular}
\medskip
\begin{flushleft}
\textbf{Notes.} Columns: (1) Richness range of the selected sub-samples (2) Selection criteria according to the abundance matching mass, $M_\text{AM}$; (3), (4) and (5)  median, 15- and 85-th percentiles of the $L_\text{GR}$ distribution in each bin, for the total sample of groups. (6), (7) and (8) same for the groups included in the $C-$sample.
\end{flushleft}
\label{table:Nlum}
\endgroup
\end{table*}

\begin{table*}
\begingroup
\setlength{\tabcolsep}{10pt} 
\renewcommand{\arraystretch}{1.5} 
\centering
\caption{Fitted parameters for the analysed galaxy groups in the whole richness range ($N_\text{GAL} \geq 1$).}
\begin{tabular}{c c | c c c | c c c}
\hline
\hline
\rule{0pt}{1.05em}%
Redshift  & $\log{M_\text{AM}}$ & \multicolumn{3}{c|}{Total sample} & \multicolumn{3}{c}{$C-$sample} \\
selection       &                                   & $L_{50}$ & $L_{15}$ & $L_{85}$ & $L_{50}$ & $L_{15}$ & $L_{85}$  \\
\hline
\rule{0pt}{1.05em}   
$z < 0.13$ & $[12.5,13.0)$ & 10.41 & 10.32 & 10.52 & 10.43 & 10.33 & 10.54 \\
           & $[13.0,13.3)$ & 10.67 & 10.61 & 10.73 & 10.67 & 10.61 & 10.74 \\
           & $[13.3,13.6)$ & 10.85 & 10.79 & 10.92 & 10.85 & 10.79 & 10.92 \\
           & $[13.6,13.9)$ & 11.04 & 10.98 & 11.13 & 11.04 & 10.98 & 11.13 \\
           & $[13.9,14.2)$ & 11.26 & 11.20 & 11.36 & 11.27 & 11.21 & 11.35 \\
           & $[14.2,14.5)$ & 11.51 & 11.43 & 11.61 & 11.51 & 11.43 & 11.61 \\
           & $[14.5,15.0)$ & 11.82 & 11.70 & 11.92 & 11.87 & 11.82 & 11.95 \\                   
\hline
$z \geq 0.13$ & $[12.5,13.0)$ & 10.43 & 10.33 & 10.53 & 10.45 & 10.35 & 10.54 \\
              & $[13.0,13.3)$ & 10.66 & 10.61 & 10.73 & 10.67 & 10.61 & 10.74 \\
              & $[13.3,13.6)$ & 10.84 & 10.79 & 10.92 & 10.84 & 10.79 & 10.92 \\
              & $[13.6,13.9)$ & 11.04 & 10.98 & 11.12 & 11.04 & 10.98 & 11.11 \\
              & $[13.9,14.2)$ & 11.23 & 11.20 & 11.31 & 11.23 & 11.20 & 11.29 \\
              & $[14.2,14.5)$ & 11.49 & 11.43 & 11.55 & 11.48 & 11.43 & 11.55 \\
              & $[14.5,15.0)$ & 11.96 & 11.89 & 12.02 & 11.96 & 11.89 & 12.02 \\
\hline         
\end{tabular}
\medskip
\begin{flushleft}
\textbf{Notes.} Columns: (1) Redshift range of the seleceted subsamples (2) Selection criteria according to the abundance matching mass, $M_\text{AM}$; (3), (4) and (5)  median, 15- and 85-th percentiles of the $L_\text{GR}$ distribution in each bin, for the total sample of groups. (6), (7) and (8) same for the groups included in the $C-$sample.
\end{flushleft}
\label{table:zlum}
\endgroup
\end{table*}

\bsp	
\label{lastpage}
\end{document}